\documentclass[twocolumn,twocolappendix]{aastex702}

\usepackage{amsmath}
\usepackage{CJK}


\newcommand{\kms}{\ensuremath{\mathrm{km\,s^{-1}}}}
\newcommand{\masyr}{\ensuremath{\mathrm{mas\,yr^{-1}}}}
\shorttitle{Stellar Kinematics toward NGC 6357}
\shortauthors{Goh et al.}

\begin{document}
\title{Stellar Kinematics and Molecular Gas toward NGC 6357: An Overlooked Earlier Stellar Generation}

\correspondingauthor{GOH TUNG WEI}

\author[0009-0007-5235-2250,gname='TUNG WEI',sname='GOH']{GOH TUNG WEI}
\affiliation{Nagoya University Affiliated Upper Secondary School, Nagoya, Aichi 464-8601, Japan}
\email{tungweigoh@gmail.com}

\author[0009-0001-8036-7296,gname='Satoya',sname='Nakano']{Satoya Nakano \begin{CJK*}{UTF8}{min}\upshape(中野 覚矢)\end{CJK*}}
\affiliation{Independent Researcher, Japan}
\email{nakanosatoya@gmail.com}

\author[0000-0002-1411-5410,gname='Kengo',sname='Tachihara']{Kengo Tachihara \begin{CJK*}{UTF8}{min}\upshape(立原 研悟)\end{CJK*}}
\affiliation{Nagoya University, Nagoya, Aichi 464-8602, Japan}
\email{k.tachihara@a.phys.nagoya-u.ac.jp}

\author[0009-0002-5520-4708,gname='Tohru',sname='Ohba']{Tohru Ohba \begin{CJK*}{UTF8}{min}\upshape(大羽 徹)\end{CJK*}}
\affiliation{Nagoya University Affiliated Upper Secondary School, Nagoya, Aichi 464-8601, Japan}
\email{oba.toru.a3@f.mail.nagoya-u.ac.jp}

\begin{abstract}
We investigate past star formation in NGC~6357 using Gaia DR3
astrometry and Three-mm Ultimate Mopra Milky Way Survey (ThrUMMS)
$^{12}$CO($J=1$--0) data. A Hierarchical Density-Based Spatial Clustering
of Applications with Noise (HDBSCAN) search
independently recovers the previously cataloged cluster G352.8+0.6 in
addition to Pismis~24, G353.1+0.6, and G353.2+0.6. Isochrone fitting
yields an age of $7.9^{+1.5}_{-1.4}$~Myr for G352.8+0.6, compared with
approximately 1--3~Myr for the three young clusters. Assuming constant
relative plane-of-sky motion, a linear traceback places G352.8+0.6
closest to their reference center approximately 1.8~Myr ago, comparable
to their inferred ages. The $+2~\kms$ CO component overlaps G352.8+0.6
in projection and shows localized depressions in integrated intensity
around three spectroscopically identified massive stars belonging to G352.8+0.6. The $-4$ and
$+2~\kms$ components show complementary projected distributions and
intermediate-velocity CO emission in position--velocity space.
Assuming that the $+2~\kms$ component was physically associated with
G352.8+0.6 and shared its transverse motion, our principal interpretation
is that the two molecular structures interacted around the epoch indicated
by the stellar traceback, approximately 1.8~Myr ago. The resulting gas
compression may have contributed to the recent episode of clustered and
high-mass star formation in NGC~6357 and to the formation of Shell~A,
which subsequent stellar feedback may have reshaped.
\end{abstract}

\section{Introduction}\label{sec:intro}

Massive-star-forming regions evolve through the coupled formation and
feedback of massive stars and the surrounding interstellar medium.
Ionizing radiation, stellar winds, and supernova explosions can disperse
molecular material or compress it into dense structures in which further
stars may form \citep{Krumholz2014,Dale2015}. Reconstructing how such
regions develop therefore requires combining stellar ages and kinematics
with the present-day spatial and velocity structure of the surrounding
molecular gas.

\begin{figure*}[!tp]
\centering
\includegraphics[width=\textwidth]{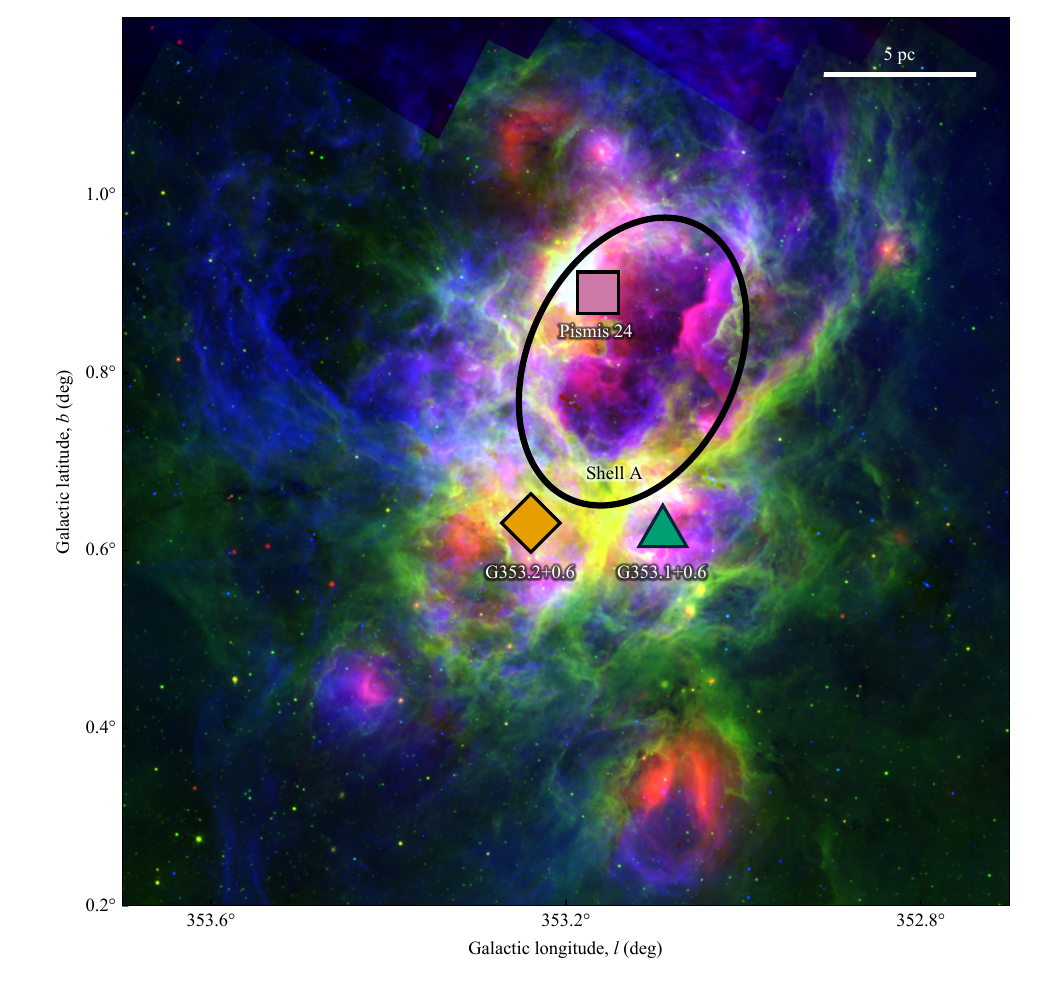}
\caption{
Three-color view of NGC~6357 in Galactic coordinates. Red shows
Spitzer/MIPSGAL $24~\micron$ emission from warm dust \citep{Carey2009},
green shows Spitzer/GLIMPSE $8~\micron$ emission \citep{Benjamin2003,Churchwell2009},
and blue shows H$\alpha$ emission from ionized gas in the VST Photometric
H$\alpha$ Survey of the Southern Galactic Plane and Bulge (VPHAS+)
\citep{Drew2014}. The $8~\micron$-bright structures surround regions
bright in H$\alpha$ and at $24~\micron$. The magenta square, green
triangle, and orange diamond mark the centers of Pismis~24, G353.1+0.6,
and G353.2+0.6, respectively. The black ellipse outlines Shell~A,
following the morphology described by \citet{Cappa2011}.
The white scale bar represents approximately 5~pc.
}
\label{fig:overview}
\end{figure*}

NGC~6357 is a Galactic massive star-forming complex at approximately
$1.7$~kpc, consistent with the mean distance of the larger G352
NGC~6334--NGC~6357 molecular-cloud complex \citep{Povich2026}.
It contains numerous massive stars,
H~{\sc ii} regions, young stellar objects (YSOs), and several stellar
clusters. Three prominent young clusters are Pismis~24, G353.1+0.6,
and G353.2+0.6. Literature identifications of G353.1+0.6 include
AH03J1725--34.4 and ESO~392-SC~11 \citep{Ahumada2007,Lima2014}.
It is associated with the broader cluster~A region of \citet{Massi2015},
which includes the Acore and AS substructures; G353.2+0.6 corresponds
to their cluster~B.
These clusters have characteristic ages of approximately $1$--$3$~Myr,
and Pismis~24 contains a particularly rich massive-star population
\citep{Wang2007,Fang2012,Russeil2017,Massi2015}.
The region also contains Shell~A, the molecular counterpart of
G353.12+0.86 near Pismis~24. \citet{Cappa2011} interpreted this shell
as material swept up by winds from the massive stars in Pismis~24.
Pismis~24 lies near the projected rim rather than the center of the
shell. Here we examine whether molecular-cloud interaction could
also have contributed to its formation.
Figure~\ref{fig:overview} shows the region and the three young clusters.

Previous studies have established that star formation in NGC~6357 is
spatially and temporally structured. X-ray and infrared observations have
identified large populations of pre-main-sequence stars, disk-bearing YSOs,
and embedded populations \citep{Wang2007,Fang2012,Russeil2020}.
Spectroscopic studies have characterized the massive-star population and
provided evidence for non-coeval stellar populations
\citep{Russeil2017}. Near-infrared astrometric studies have also examined
the internal kinematic structure of NGC~6357 and reported asymmetric
proper-motion distributions among its stellar populations
\citep{OrdenesHuanca2024}. More recently, \citet{Povich2026} combined
X-ray- and infrared-selected young stars with Gaia DR3 astrometry to
investigate the three-dimensional structure and large-scale kinematics of
the young population in the broader NGC~6334--NGC~6357 complex.

Despite this progress, much of the stellar census of NGC~6357 has relied
on diagnostics that preferentially identify young stars. Infrared excess declines as circumstellar disks dissipate
\citep{Haisch2001}, while stellar X-ray properties also
evolve with age and stellar mass
\citep{Preibisch2005,Getman2022}.
Populations older than several Myr can therefore
be less conspicuous in youth-selected samples. Identifying such clusters
and measuring their ages and motions provides an additional way to trace
past star formation that is difficult to recover from the youngest
populations alone.

Gaia positions, parallaxes, and proper motions provide a complementary
approach by allowing stellar groups to be identified through astrometric
coherence without requiring conventional youth indicators. Density-based
searches in Gaia astrometric space have recovered known clusters and
identified additional stellar aggregates
\citep{HuntReffert2021,HuntReffert2023}. Of particular relevance here,
the stellar aggregate near $(l,b)\simeq(352.8^\circ,0.6^\circ)$ has
already been cataloged as PHOC~17 and UFMG~76
\citep{HuntReffert2021,Ferreira2021}. \citet{Ferreira2021} estimated
an age of order 8~Myr for UFMG~76, substantially older than the principal
young clusters in NGC~6357. Its possible dynamical relationship with the
younger stellar populations in NGC~6357, however, remains to be
investigated.

Molecular-gas observations have likewise motivated dynamical formation
scenarios for NGC~6357. In particular, \citet{Fukui2018} proposed a
cloud--cloud collision involving the main molecular component near
$-4~\kms$ and a blueshifted component near $-15~\kms$, based on their
spatial and velocity structure. In contrast, we examine the $-4~\kms$
component together with a distinct component near $+2~\kms$ and relate
their spatial and velocity structure to the independently measured
stellar motion of G352.8+0.6. This provides a different test of whether
stellar kinematics and molecular-gas structure can be linked within the
formation process of NGC~6357.

We therefore perform an unsupervised Gaia DR3 cluster search around
NGC~6357, independently recovering G352.8+0.6 together with the three
young clusters. We estimate the clusters' distances, ages, extinctions,
and total initial masses and trace the relative plane-of-sky motion of
G352.8+0.6 backward under a constant-velocity approximation. We then
combine the stellar kinematics with the Three-mm Ultimate Mopra
Milky Way Survey (ThrUMMS)
$^{12}$CO($J=1$--0) observations \citep{Barnes2015} to examine the
relationship between G352.8+0.6 and the $+2~\kms$ molecular gas and the
connection between the $-4$ and $+2~\kms$ velocity structures. Through
these measurements, we investigate their possible implications for the
formation of the young clusters and the formation and subsequent shaping
of Shell~A.

\section{Data}\label{sec:data}

\subsection{Gaia DR3}\label{subsec:data-gaia}

We selected Gaia DR3 sources \citep{2016A&A...595A...1G,GaiaDR3} within $3^\circ$ of the
adopted NGC~6357 center,
$(\alpha,\delta)=(261.61^\circ,-34.22^\circ)$
\citep{RamirezTannus2020}. We required positions, parallax, both
proper-motion components, and photometry in the $G$, $G_{\rm BP}$, and
$G_{\rm RP}$ bands, yielding 4,727,518 sources. Cluster identification
used the five astrometric quantities
$(\alpha,\delta,\mu_{\alpha *},\mu_\delta,\varpi)$, where
$\mu_{\alpha *}=\mu_\alpha\cos\delta$. No additional astrometric-quality
cut, including a RUWE cut, was applied before clustering. The astrometric
reference epoch is J2016.0.

Clustering used raw DR3 parallaxes. After membership identification, we
applied the source-by-source zero-point correction of
\citet{Lindegren2021} for candidate screening and cluster-parameter
estimation. Quoted cluster parallaxes and distances use these corrected
values unless stated otherwise. The Gaia $G$, $G_{\rm BP}$, and
$G_{\rm RP}$ measurements were subsequently used for the
color--magnitude analysis described in
Section~\ref{subsec:age-extinction}.

We used \texttt{gaiadr3-zeropoint} version 0.1.0, evaluating the
zero point from $G$, ecliptic latitude, astrometric solution type, and
\texttt{nu\_eff\_used\_in\_astrometry} for five-parameter solutions or
\texttt{pseudocolour} for six-parameter solutions. The correction is
$\varpi_{\rm corr}=\varpi_{\rm DR3}-\varpi_{\rm zp}$. All 257 members
of the four principal clusters have the required inputs. The default
implementation's boundary treatment was retained for sources outside
the published calibration ranges. There are 55 such members in total: 29, 8,
8, and 10 in Pismis~24, G353.1+0.6, G353.2+0.6, and G352.8+0.6,
respectively.
The effect of restricting the sample to the calibrated ranges is
examined in Appendix~\ref{app:parallax}.

\subsection{VIRAC2}\label{subsec:data-virac2}

We used the second VVV Infrared Astrometric Catalogue (VIRAC2;
\citealt{Smith2025}), based on VVV/VVVX observations, to compare optical
and near-infrared cluster recovery. Within the same $3^\circ$ field, we
required positions, both proper-motion components, and uncertainties
smaller than $2~\masyr$ in each component, yielding 17,511,568 sources.
VIRAC2 positions are referenced to J2014.0, whereas Gaia DR3 positions are
referenced to J2016.0.

The typical VIRAC2 parallax uncertainty, approximately $1$~mas for
$11<K_s<14$, exceeds the expected $\sim0.6$~mas parallax of NGC~6357;
the corresponding proper-motion uncertainty is approximately
$0.37~\masyr$ per component \citep{Smith2025}. We therefore clustered
only in $(\alpha,\delta,\mu_{\alpha *},\mu_\delta)$.
The cross-match of the Gaia-selected members with VIRAC2 is described
in Appendix~\ref{app:virac2-crossmatch}. The comparison of cluster
recovery, catalog source density, and astrometric precision is discussed
in Appendix~\ref{app:virac2}.

\subsection{ThrUMMS Molecular-line Data}
\label{subsec:data-thrumms}

We used ThrUMMS $^{12}$CO($J=1$--0) data at 115.271~GHz
\citep{Barnes2015} to examine the molecular-gas distribution and
line-of-sight velocity structure toward NGC~6357. The data have an
effective angular resolution of approximately $72\arcsec$, a pixel size
of $24\arcsec$, and a velocity-channel spacing of $0.35048~\kms$.
Velocities are expressed with respect to the kinematic local standard
of rest, $V_{\rm LSR}$.

The $^{12}$CO data used here are calibrated on the ThrUMMS
extended-source temperature scale,
\begin{equation}
T_r^*=\frac{T_A^*}{\eta_c},
\end{equation}
with $\eta_c=0.55$ for $^{12}$CO \citep{Barnes2015}. We therefore
applied no additional efficiency correction.

A robust estimate from line-free velocity channels gives a per-channel
rms noise of $1.414$~K over
$352.2^\circ\leq l\leq353.5^\circ$ and
$0.0^\circ\leq b\leq1.0^\circ$.

The CO integrated-intensity maps were calculated directly from the data
without additional spatial smoothing, reprojection, or per-pixel
baseline subtraction. Pixels containing invalid values within a selected
velocity interval were excluded from the corresponding integration.
The adopted velocity intervals and the propagation of the channel noise
into the integrated-intensity maps are described in
Section~\ref{subsec:gas-analysis}.

\subsection{Ancillary Infrared and Spectroscopic Data}
\label{subsec:data-ancillary}

Spitzer/GLIMPSE $8~\micron$ images \citep{Benjamin2003,Churchwell2009} provide the
background for the CO maps. The overview in Figure~\ref{fig:overview}
also uses Spitzer/MIPSGAL $24~\micron$ \citep{Carey2009} and VPHAS+
H$\alpha$ imaging \citep{Drew2014} to show warm dust and ionized gas.
These images were used for qualitative morphological comparisons.

We obtained Gaia DR3 radial velocities \citep{Katz2023} for the selected
cluster members. For the VLT/KMOS observations described by
\citet{RamirezTannus2020}, we adopted the radial velocities and reported
uncertainties in Table B.2 of \citet{RamirezTannus2021}. These sources
were associated with Gaia members using source identifiers and
proper-motion-propagated positions. The KMOS cross-match criteria,
individual measurements, velocity reference frames, and reliability
assessment are described in Appendix~\ref{app:rv}.

\section{Analysis}\label{sec:analysis}

\subsection{Stellar Cluster Identification with HDBSCAN}
\label{subsec:hdbscan}

We applied Hierarchical Density-Based Spatial Clustering of Applications
with Noise (HDBSCAN), a density-based algorithm that identifies groups
against a lower-density background \citep{Campello2015,McInnes2017}, to
the five-dimensional Gaia sample of positions, proper motions, and
parallaxes. Each coordinate was centered on its mean and scaled by its
standard deviation over the full input sample. All sources assigned to
a group rather than to the algorithm's noise population were retained,
without an additional membership-strength threshold.

Candidates were inspected if at least one member lies within
$352.6^\circ\leq l\leq353.6^\circ$ and
$0.2^\circ\leq b\leq1.2^\circ$; this criterion applies to member
positions rather than candidate centers. For the principal analysis,
we retained candidates with median corrected parallaxes satisfying
$0.48<\tilde{\varpi}_{\rm corr}<0.67$~mas. This practical screening
window was motivated by the NGC~6334--NGC~6357 parallax range of
\citet{Russeil2020}, whose criterion used uncorrected Gaia DR2
parallaxes. We then assessed entire candidates using their spatial
distributions, proper-motion coherence, and color--magnitude diagrams
(CMDs), without removing individual members according to CMD position.
Additional candidates without a parallax restriction are listed in
Appendix~\ref{app:candidates}; the algorithm settings are given in
Appendix~\ref{app:analysis-hdbscan}.

For a comparison with VIRAC2, we also searched both catalogs using only
positions and proper motions, omitting parallax and standardizing each
catalog independently. The comparison and its settings are described in
Appendix~\ref{app:virac2}.

\subsection{Cluster Distances, Ages, and Extinction}
\label{subsec:age-extinction}

For each cluster, we adopted the median zero-point-corrected member
parallax, $\tilde{\varpi}_{\rm corr}$, and estimated its distance as
\begin{equation}
d~[\mathrm{kpc}]
=\frac{1}{\tilde{\varpi}_{\rm corr}~[\mathrm{mas}]}.
\label{eq:cluster-distance}
\end{equation}
The 16th--84th percentile intervals were obtained by bootstrap
resampling, drawing with replacement from each fixed member list and
recomputing the distance. They describe member-resampling uncertainty
and exclude separately propagated parallax measurement errors and
residual Gaia systematics. The latter are considered in
Section~\ref{subsubsec:distance} and Appendix~\ref{app:parallax}.

Cluster centers were defined by median member right ascension and
declination. For angular separation $\theta_i$ from this center, the
projected radius is $d\tan\theta_i$; $r_{90}$ is its 90th percentile
over the selected members.

We fitted Gaia $G$ versus $G_{\rm BP}-G_{\rm RP}$ CMDs with PARSEC
isochrones, which describe stellar populations of a common age
\citep{Bressan2012,Chen2014,Chen2015,Tang2014,Marigo2017,
Pastorelli2019,Pastorelli2020}. We fixed the distance to the value
above and varied age and extinction, adopting $Z_{\rm ini}=0.0152$
and the extinction law of \citet{Fitzpatrick1999} with $R_0=3.1$.
Extinction in each Gaia band depends on stellar temperature. The fitted
parameter is monochromatic extinction $A_0$, reported as $A_V$ under
the approximation $A_V\simeq A_0$.

For each cluster, the magnitude and color axes were standardized using
the observed member means and standard deviations. All selected
members were retained. If $r_i$ is a member's distance to the nearest
point on a trial isochrone in this standardized CMD, the best-fitting
curve minimizes
\begin{equation}
R=\sum_i r_i.
\label{eq:isochrone-residual}
\end{equation}
We also assigned empirical likelihood weights using the residual
dispersion around each trial curve, allowing the observed scatter to
represent differential extinction and other unmodeled variations.
With uniform grid weighting in log age and $A_0$, we marginalized over
the other parameter and report the medians and 16th--84th percentiles.
These intervals exclude systematic uncertainties in distance, stellar
models, the extinction prescription, and other unmodeled scatter.
The model settings, likelihood definition, and numerical details are
given in Appendix~\ref{app:analysis-cmd}.

\subsection{Total Initial Cluster Mass Estimates}
\label{subsec:mass}

We assigned each member an initial stellar mass from the nearest point
on its cluster's PARSEC isochrone in the standardized CMD, using the
age, extinction, and fixed distance described in
Section~\ref{subsec:age-extinction}. All available evolutionary phases
were included, without an additional member-residual cut. We estimated
an empirical completeness threshold from the member $G$-magnitude
distribution and converted it to an initial-mass threshold.

We normalized the Kroupa initial mass function (IMF;
\citealt{Kroupa2001}), which describes the distribution of stellar
birth masses, to the selected members above this threshold. The
normalization upper limit was the smaller of $100~M_\odot$ and the
largest initial mass still represented by the adopted isochrone at the
cluster age. Integrating the normalized IMF over
$0.08$--$100~M_\odot$ then gives the approximate total initial stellar
mass, including low-mass stars below the threshold and high-mass stars
that have completed their evolution. No additional corrections were
made for HDBSCAN incompleteness, unresolved multiplicity, or dynamically
lost members. The thresholds, model settings, and normalization
equations are given in Appendix~\ref{app:analysis-mass}.

\subsection{Plane-of-sky Cluster Kinematics and Linear Traceback Analysis}
\label{subsec:traceback}

We compared positions and proper motions on a common plane tangent to
the sky. Cluster centers were the median member right ascension and
declination, and each cluster's representative motion was the
unweighted mean of its members' projected proper motions. The
reference center and motion were the equal-weight means of the three
young-cluster centers and motions, independent of their member counts.
The projection details are given in Appendix~\ref{app:analysis-traceback}.

For G352.8+0.6, we defined
$\Delta\boldsymbol{\mu}=\boldsymbol{\mu}_{\rm cl}-\boldsymbol{\mu}_{\rm ref}$
and assumed constant relative plane-of-sky motion. Its displacement at
lookback time $\tau>0$ is
\begin{equation}
\boldsymbol{\theta}(\tau)
=\boldsymbol{\theta}_0-\Delta\boldsymbol{\mu}\,\tau,
\label{eq:angular-traceback}
\end{equation}
where $\boldsymbol{\theta}_0$ is the present displacement from the
reference center. The epoch of minimum angular separation over
$\tau\geq0$ is
\begin{equation}
\tau_{\rm min}
=\max\left[0,
\frac{\boldsymbol{\theta}_0\cdot\Delta\boldsymbol{\mu}}
{\left|\Delta\boldsymbol{\mu}\right|^2}\right].
\label{eq:closest-time}
\end{equation}

The traceback itself is independent of distance. We adopted
$d_{\rm ref}=1.67$~kpc, the mean distance of the G352 molecular-cloud
complex \citep{Povich2026}, only to convert angular separations and
proper motions to projected physical scales. These are not
three-dimensional separations or velocities.

We assessed sensitivity to member sampling by bootstrap resampling the
four fixed member lists and recalculating their centers and motions.
This assesses member-resampling variation rather than the full
uncertainty from astrometric errors, residual systematics, and the
constant-velocity assumption. The implementation is described in
Appendix~\ref{app:analysis-traceback}. Reliable cluster systemic radial
velocities could not be determined from the available measurements
(Appendix~\ref{app:rv}), so the traceback is restricted to the plane
of the sky.

\subsection{Molecular Gas Analysis}
\label{subsec:gas-analysis}

We fitted three Gaussian components and a linear baseline to the mean
CO spectrum to characterize its principal peaks. The baseline was used
only in this fit; integrated-intensity maps were calculated directly
from the CO cube. We adopted
$-8.00\leq V_{\rm LSR}<-0.31~\kms$ and
$-0.31\leq V_{\rm LSR}<+4.25~\kms$ for the $-4$ and $+2~\kms$
components, respectively, placing the interval limits at the nearest
available velocity channels. Integrated intensity and its statistical
uncertainty were calculated as
\begin{equation}
W_{\rm CO}(l,b)
=\int_{V_1}^{V_2}T_r^\ast(l,b,V_{\rm LSR})\,dV_{\rm LSR},
\label{eq:wco}
\end{equation}
\begin{equation}
\sigma_W=\sigma_T\,\Delta v\,\sqrt{N_{\rm ch}},
\label{eq:wco-noise}
\end{equation}
where $\sigma_T$ is the per-channel rms, $\Delta v$ is the channel
spacing, and $N_{\rm ch}$ is the number of channels, assumed
independent. CO distributions were compared with the Spitzer
$8~\micron$ image, cluster positions, and Shell~A.

We extracted a position--velocity (PV) diagram by averaging CO
temperatures across a strip aligned with the relative-motion direction
of G352.8+0.6, with position referenced to its center. The averaging
region, noise values, and extraction settings are given in
Appendix~\ref{app:analysis-gas}. The wide-field comparison including
NGC~6334 is presented in Appendix~\ref{app:extended-co}.

\section{Results}\label{sec:results}

\subsection{Cluster Identification and Physical Properties}
\label{subsec:results-clusters}

The five-dimensional Gaia search identified 3733 candidate overdensities
over the full field and recovered Pismis~24, G353.1+0.6, G353.2+0.6,
and G352.8+0.6. The last corresponds to the cataloged PHOC~17/UFMG~76
cluster \citep{HuntReffert2021,Ferreira2021}. Their selected members
are shown in Figure~\ref{fig:cluster-map}; the other candidates and
target selection are described in Appendix~\ref{app:candidates}.

\begin{figure*}[!tp]
\centering
\includegraphics[width=\textwidth]{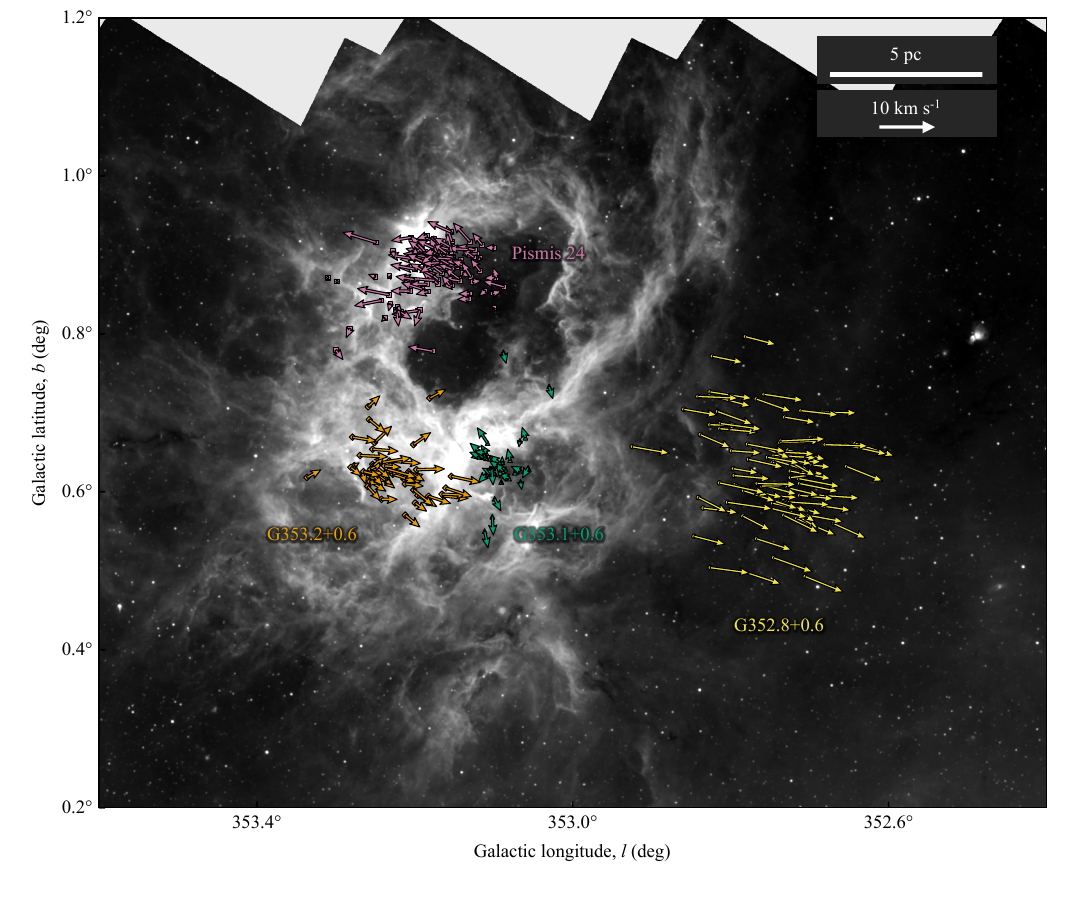}
\caption{
Selected Gaia DR3 members of the four principal clusters overlaid
on the Spitzer/GLIMPSE $8~\micron$ image \citep{Benjamin2003,Churchwell2009}.
Magenta squares, green triangles, orange diamonds, and yellow circles
represent individual members of Pismis~24, G353.1+0.6, G353.2+0.6,
and G352.8+0.6, respectively. Arrows in the corresponding colors show
individual stellar proper motions relative to the equal-weight mean
motion of the three young clusters, defined in
Section~\ref{subsec:traceback}. The arrow key corresponds to a
transverse speed of approximately $10~\kms$, and the white scale bar
represents approximately 5~pc.
}
\label{fig:cluster-map}
\end{figure*}

\begin{figure}[!tp]
\centering
\includegraphics[width=\columnwidth]{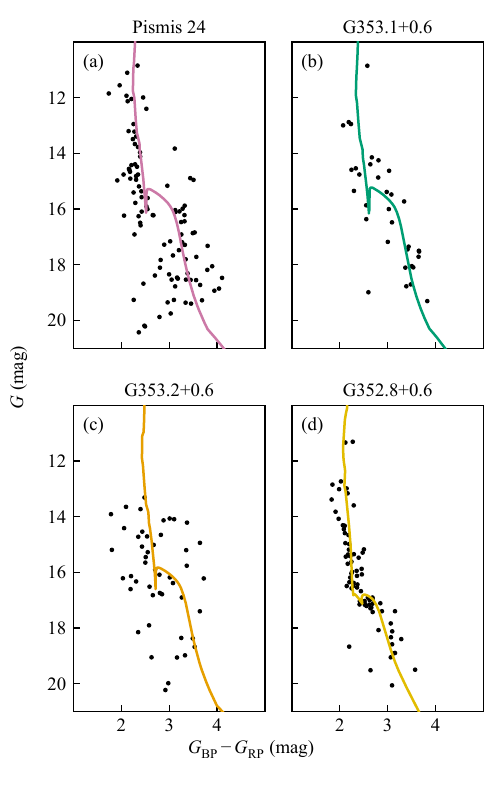}
\caption{
Gaia color--magnitude diagrams of (a) Pismis~24,
(b) G353.1+0.6, (c) G353.2+0.6, and (d) G352.8+0.6.
The horizontal and vertical axes show the observed
$G_{\rm BP}-G_{\rm RP}$ color and apparent $G$ magnitude, respectively.
Black points are the selected Gaia members. The colored solid curves
are the best-fit PARSEC isochrones, selected by minimizing the summed
CMD residual $R=\sum_i r_i$ defined in
Section~\ref{subsec:age-extinction}. Distance and extinction are
included in the model curves. Table~\ref{tab:cluster-properties}
separately reports the estimated ages and extinctions as the medians
of the marginalized empirical-likelihood distributions, with their
16th--84th percentile intervals.
}
\label{fig:cluster-cmds}
\end{figure}

Table~\ref{tab:cluster-properties} summarizes the four clusters.
The inferred distances of all four clusters lie near the mean
heliocentric distance of the G352 molecular-cloud complex,
$1.67\pm0.08$~kpc \citep{Povich2026}. The quoted spread describes
the large-scale complex rather than an uncertainty on the distance
of NGC~6357 alone. G352.8+0.6 has the smallest
nominal distance, lying approximately $0.05$--$0.13$~kpc closer than the
three young clusters; the possible significance of this offset is
discussed in Section~\ref{subsubsec:distance}.
The CMD fits in Figure~\ref{fig:cluster-cmds} give an age of
$7.9^{+1.5}_{-1.4}$~Myr for G352.8+0.6, compared with approximately
$1$--$3$~Myr for the young clusters. Its fitted extinction,
$A_V\simeq4.82^{+0.06}_{-0.07}$~mag under the $A_V\simeq A_0$
approximation, is also lower than their median values of approximately
$5.25$--$5.55$~mag. G352.8+0.6 has 76 selected members and the
largest $r_{90}$, approximately 3.5~pc.

Figure~\ref{fig:cluster-cmds} shows the curves minimizing $R$, whereas
Table~\ref{tab:cluster-properties} reports the marginalized medians.

The IMF-extrapolated total initial masses are approximately 1017, 388,
690, and $715~M_\odot$ for Pismis~24, G353.1+0.6, G353.2+0.6,
and G352.8+0.6, respectively. The Pismis~24 estimate is provisional:
six members were assigned to post-main-sequence model points with
initial masses above $100~M_\odot$ and fall outside the adopted IMF
normalization interval (Appendix~\ref{app:analysis-mass}).

G352.8+0.6 was recovered in the four-dimensional Gaia search but not as
a distinct VIRAC2 group, although 73 of its 76 Gaia-selected members
have VIRAC2 counterparts. Its relatively diffuse distribution and the
much denser VIRAC2 field plausibly reduce its clustering contrast:
within the same aperture, VIRAC2 contains approximately 30 times as
many catalog sources as Gaia. Thus, the non-recovery concerns a
distinct stellar grouping rather than the absence of its individual
stars from the catalog. The comparison and possible contributing
factors are discussed in Appendix~\ref{app:virac2}.

\begin{deluxetable*}{lccccccc}
\tablecaption{
Estimated Properties of the Four Principal Clusters
\label{tab:cluster-properties}
}
\tablehead{
\colhead{Cluster} & \colhead{$N$} & \colhead{$d$} &
\colhead{$r_{90}$} & \colhead{Age} & \colhead{$A_V\simeq A_0$} &
\colhead{$M_{\rm ini}$} & \colhead{$(l,b)$} \\
\colhead{} & \colhead{} & \colhead{(kpc)} & \colhead{(pc)} &
\colhead{(Myr)} & \colhead{(mag)} & \colhead{($M_\odot$)} &
\colhead{(deg)}
}
\startdata
Pismis~24 & 101 & $1.68^{+0.01}_{-0.02}$ & 2.49
& $2.4^{+0.7}_{-1.0}$ & $5.25^{+0.09}_{-0.09}$
& 1017\tablenotemark{a} & (353.17, 0.89) \\
G353.1+0.6 & 33 & $1.71^{+0.01}_{-0.05}$ & 1.83
& $1.13^{+0.10}_{-0.09}$\tablenotemark{b} & $5.44^{+0.15}_{-0.13}$
& 388 & (353.09, 0.63) \\
G353.2+0.6 & 47 & $1.63^{+0.04}_{-0.01}$ & 2.38
& $2.5^{+1.9}_{-0.9}$ & $5.55^{+0.13}_{-0.46}$
& 690 & (353.24, 0.63) \\
G352.8+0.6 & 76 & $1.58^{+0.02}_{-0.02}$ & 3.52
& $7.9^{+1.5}_{-1.4}$ & $4.82^{+0.06}_{-0.07}$
& 715 & (352.75, 0.63)
\enddata
\tablenotetext{a}{The Pismis~24 mass is provisional pending resolution of
six model assignments with initial masses above $100~M_\odot$; see
Section~\ref{subsec:results-clusters} and Appendix~\ref{app:analysis-mass}.}
\tablenotetext{b}{The G353.1+0.6 age remains provisional because the
fitting distribution approaches the lower age-grid boundary. Its mass
estimate is also conditional on the adopted age.}
\tablecomments{
$N$ is the number of selected Gaia members. Distances are the
reciprocals of the median zero-point-corrected member parallaxes;
their intervals are the 16th--84th percentiles of member-resampling
bootstraps and exclude separately propagated measurement errors and
residual Gaia systematics. $r_{90}$ is the projected radius enclosing
90\% of the selected members at the tabulated distance.
The coordinates $(l,b)$ are the median RA/Dec centers transformed to
Galactic coordinates. Ages and extinctions are estimated as the
medians of the marginalized empirical-likelihood distributions at
fixed CMD distances; their intervals are the 16th--84th percentiles.
The best-fit curves displayed in Figure~\ref{fig:cluster-cmds} minimize
the summed CMD residual. $A_V$ denotes the approximation
$A_V\simeq A_0$, where $A_0$ is the fitted monochromatic extinction.
$M_{\rm ini}$ is the approximate total initial stellar mass obtained
by IMF extrapolation over $0.08$--$100~M_\odot$, using the adopted CMD
parameters and completeness thresholds described in
Section~\ref{subsec:mass}.
}
\end{deluxetable*}

\subsection{Plane-of-sky Kinematics and Radial-velocity Constraints}
\label{subsec:results-kinematics}

\begin{figure*}[!tp]
\centering
\includegraphics[width=\textwidth]{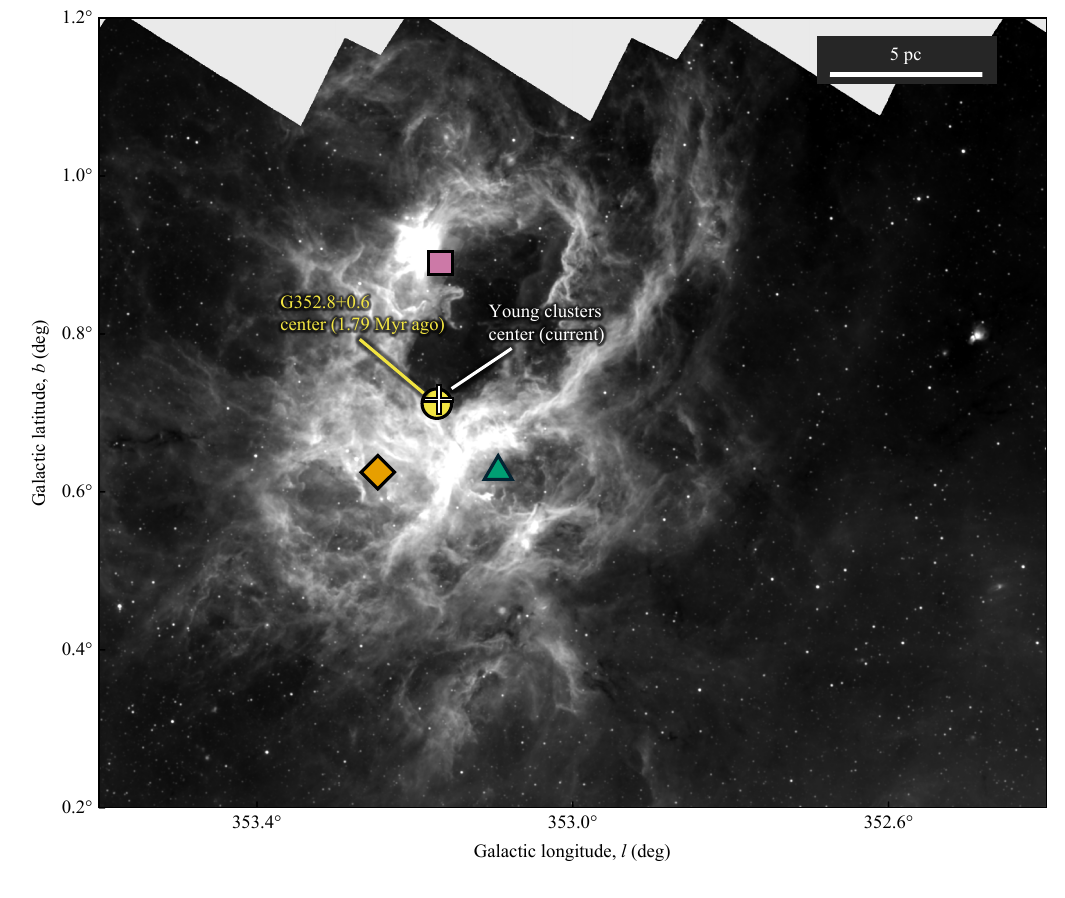}
\caption{
Back-projected center of G352.8+0.6 on the present-day
Spitzer/GLIMPSE $8~\micron$ image \citep{Benjamin2003,Churchwell2009}.
The yellow circle marks the cluster center approximately 1.8~Myr ago
(labeled 1.79~Myr in the figure), calculated by tracing its relative
proper motion backward at constant velocity as described in
Section~\ref{subsec:traceback}. The white cross marks the current
equal-weight reference center of the three young clusters. The magenta
square, green triangle, and orange diamond mark their individual current
centers: Pismis~24, G353.1+0.6, and G353.2+0.6, respectively.
The yellow and white leader lines identify the two nearly coincident
centers. This is a plane-of-sky reconstruction. The white scale bar
represents approximately 5~pc.
}
\label{fig:traceback}
\end{figure*}

The projected angular separation between G352.8+0.6 and the
young-cluster reference center is currently increasing
(Figure~\ref{fig:cluster-map}). Its relative proper motion has magnitude
approximately $0.85~\masyr$, corresponding to approximately $6.7~\kms$
at the adopted common distance of 1.67~kpc. The linear traceback
(Figure~\ref{fig:traceback}) places the closest projected approach
approximately 1.8~Myr ago; the bootstrap median also rounds to
1.8~Myr. The bootstrap assesses variation within the adopted member
lists and does not quantify the full uncertainty from astrometric
errors, residual systematics, or the constant-velocity assumption.
We therefore quote this as an approximate characteristic epoch.

Reliable cluster systemic radial velocities could not be determined from the available measurements, so the traceback remains two-dimensional. The radial-velocity measurements and their reliability assessment are presented in Appendix~\ref{app:rv}.

\subsection{Molecular Gas Distribution and Velocity Structure}
\label{subsec:results-gas}

\subsubsection{Velocity Components and Spatial Distribution}
\label{subsubsec:gas-spatial}

\begin{figure}[!tp]
\centering
\includegraphics[width=\columnwidth]{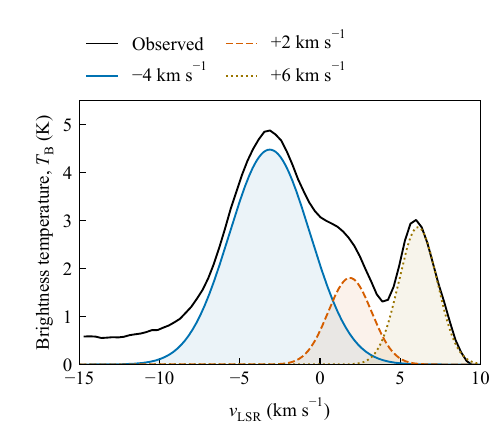}
\caption{
Spatially averaged ThrUMMS $^{12}$CO($J=1$--0) spectrum
toward NGC~6357. The horizontal axis gives $V_{\rm LSR}$, and the
vertical axis gives the mean CO temperature in kelvin. The temperature
is on the ThrUMMS $T_r^*$ scale, labeled $T_B$ in the figure.
The black curve is the observed spectrum. The blue solid, orange
dashed, and gold dotted curves show the fitted Gaussian components
referred to as the $-4$, $+2$, and $+6~\kms$ components, respectively;
the pale shading identifies the corresponding profiles.
The spatial averaging and the three-component fit, which includes a
linear baseline, are described in Section~\ref{subsec:gas-analysis}.
}
\label{fig:co-spectrum}
\end{figure}

The mean CO spectrum contains three principal features, referred to
here as the $-4$, $+2$, and $+6~\kms$ components
(Figure~\ref{fig:co-spectrum}); their fitted Gaussian centers are
$-3.17$, $+1.84$, and $+6.10~\kms$, respectively. The $+6~\kms$ emission has
previously been interpreted as foreground gas \citep{Russeil2019}; we
focus on the $-4$ and $+2~\kms$ components.

\begin{figure*}[!tp]
\centering
\includegraphics[width=\textwidth]{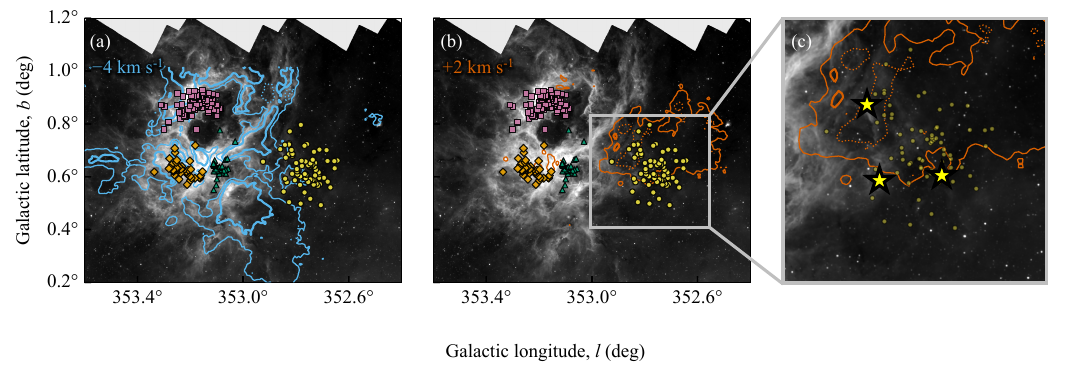}
\caption{
ThrUMMS $^{12}$CO($J=1$--0) integrated-intensity contours on
the Spitzer/GLIMPSE $8~\micron$ image \citep{Benjamin2003,Churchwell2009}.
The left panel shows the $-4~\kms$ component in blue, integrated over
$-8.00\leq V_{\rm LSR}<-0.31~\kms$; the middle panel shows the
$+2~\kms$ component in orange, integrated over
$-0.31\leq V_{\rm LSR}<+4.25~\kms$.
The gray box in the middle panel and its connecting lines identify
the enlarged region shown in the right panel. Magenta squares, green
triangles, orange diamonds, and yellow circles mark selected members
of Pismis~24, G353.1+0.6, G353.2+0.6, and G352.8+0.6, respectively.
In the right panel, the three yellow star-shaped symbols mark the
massive stars classified spectroscopically by \citet{Russeil2017}
and assigned to G352.8+0.6 by our Gaia-based membership selection.
Thin and thick contours show $15\sigma_W$ and
$30\sigma_W$, respectively; dotted segments enclose local depressions
in the integrated intensity. For the blue component,
$\sigma_W\simeq2.32~\mathrm{K\,km\,s^{-1}}$, giving contour levels of
34.9 and 69.7~K~km~s$^{-1}$; for the orange component,
$\sigma_W\simeq1.79~\mathrm{K\,km\,s^{-1}}$, giving levels of
26.8 and 53.6~K~km~s$^{-1}$.
}
\label{fig:co-maps}
\end{figure*}

The $-4~\kms$ emission is distributed broadly around the three young
clusters and Shell~A, with relatively low integrated intensity in the
shell's projected interior (Figure~\ref{fig:co-maps}, left panel). The $+2~\kms$
component overlaps G352.8+0.6 and extends toward Shell~A
(Figure~\ref{fig:co-maps}, middle panel). Around three massive stars
classified spectroscopically by \citet{Russeil2017} and assigned to
G352.8+0.6 by our Gaia-based membership selection, its integrated intensity is locally
lower than in the surrounding emission (Figure~\ref{fig:co-maps}, right panel).

The $-4$ and $+2~\kms$ components show complementary projected
distributions: stronger emission in one interval tends to coincide with
weaker emission in the other (Figure~\ref{fig:co-maps}, left and middle panels).
The wide-field maps in Appendix~\ref{app:extended-co} show that emission
within the velocity interval assigned to the $+2~\kms$ component is also
prominent toward the central region of the neighboring massive
star-forming complex NGC~6334. Emission in both the $-4$ and
$+2~\kms$ intervals is distributed across the region between
NGC~6334 and NGC~6357, where the two velocity-selected distributions
also show complementary morphology in several regions.

\subsubsection{Position--Velocity Structure}
\label{subsubsec:gas-pv}

\begin{figure}[!tp]
\centering
\includegraphics[width=\columnwidth]{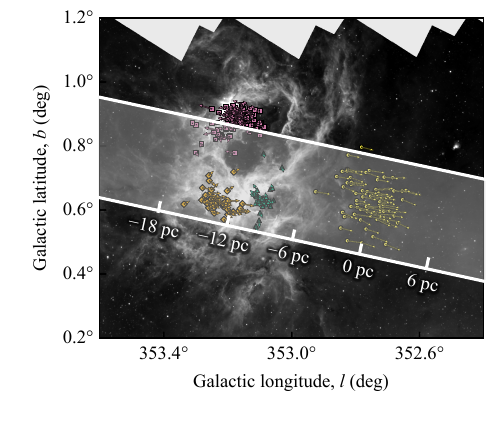}
\caption{
Sky region used to extract the position--velocity diagram in
Figure~\ref{fig:pv}. The background is the Spitzer/GLIMPSE
$8~\micron$ image \citep{Benjamin2003,Churchwell2009}; the cluster-member symbols and
relative proper-motion arrows follow Figure~\ref{fig:cluster-map}.
The gray strip between the two white lines is aligned with the
projected relative-motion direction of G352.8+0.6 and has a full
width of $17.72\arcmin$. CO temperatures were averaged across the strip
within each position bin. The ticks on its lower boundary show the
projected position coordinate in parsecs: zero is referenced to the
current center of G352.8+0.6, and positive offsets run toward decreasing
Galactic longitude and latitude. The extraction is described in
Section~\ref{subsec:gas-analysis}.
}
\label{fig:pv-region}
\end{figure}

\begin{figure*}[!tp]
\centering
\includegraphics[width=\textwidth]{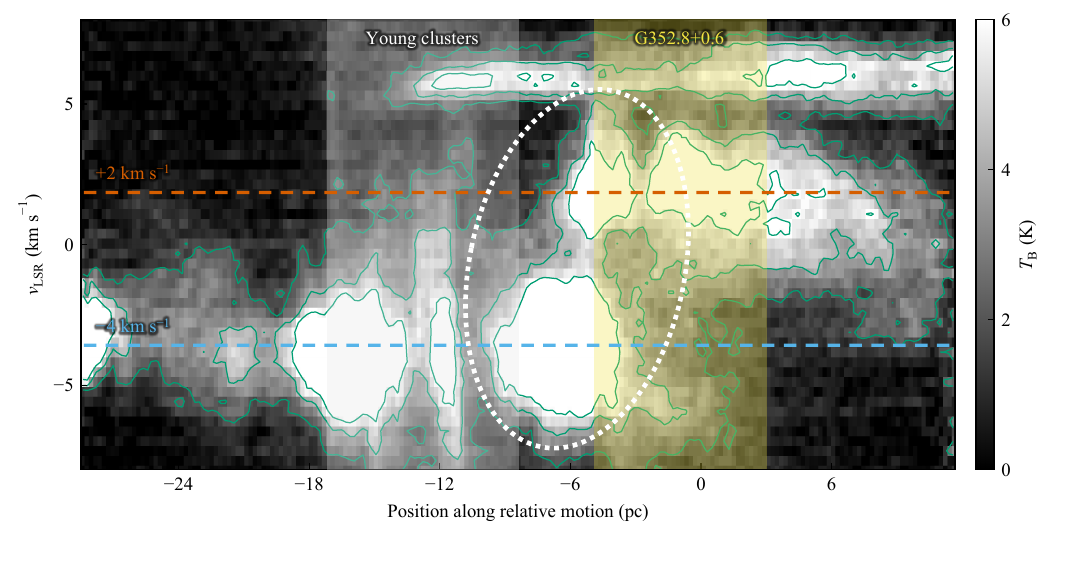}
\caption{
Position--velocity diagram of the ThrUMMS $^{12}$CO($J=1$--0)
emission averaged across the strip shown in Figure~\ref{fig:pv-region}.
The horizontal axis gives projected position along the relative-motion
direction of G352.8+0.6, with zero at its current center and positive
offsets toward decreasing Galactic longitude and latitude. The vertical
axis gives $V_{\mathrm{LSR}}$. The grayscale and color bar represent
the mean CO temperature on the ThrUMMS $T_r^{*}$ scale, labeled $T_B$
in the figure, in kelvin; lighter pixels indicate stronger emission.
Green solid contours are drawn at 2, 4, and 6 K on the same temperature
scale. The blue and orange dashed horizontal lines mark the
representative velocities of $-4$ and $+2~\kms$, respectively.
The gray and yellow vertical bands show the projected spatial extents
of the three young clusters and G352.8+0.6, respectively, along the slice.
The white dotted ellipse highlights the region of bridge-like
intermediate-velocity emission between the two components.
}
\label{fig:pv}
\end{figure*}

Figure~\ref{fig:pv-region} shows the extraction strip. The PV diagram
along the relative-motion direction of G352.8+0.6 shows
both the $-4$ and $+2~\kms$ features, with position referenced to its
center (Figure~\ref{fig:pv}). CO emission extends continuously through
intermediate velocities between the two peaks, producing a bridge-like
structure in this PV slice.

\section{Discussion}\label{sec:discussion}

\subsection{Assessing the Physical Association of G352.8+0.6 with NGC~6357}
\label{subsec:association}

G352.8+0.6 lies on the same broad distance scale as the three young
clusters and is older, less extinguished, more spatially extended, and
kinematically distinct. We assess its physical relationship with
NGC~6357 through the relative distances, the stellar kinematic
timescale, and the relationship to the $+2~\kms$ gas.

\subsubsection{Relative Cluster Distances and Parallax Systematics}
\label{subsubsec:distance}

The inferred distances of all four clusters lie near the
$1.67\pm0.08$~kpc mean distance of the G352 molecular-cloud
complex \citep{Povich2026}. The nominal
values place G352.8+0.6 approximately 50--130~pc in front of the three
young clusters.

Residual Gaia parallax systematics remain after source-by-source
zero-point correction \citep{Lindegren2021,VasilievBaumgardt2021}.
The sensitivity analysis in Appendix~\ref{app:parallax} shows that
differential parallax offsets of the tested magnitude can substantially
reduce the nominal relative parallax differences. The relative
line-of-sight placement is therefore sensitive to differential
systematics. The distance measurements therefore do not exclude a
physical association between G352.8+0.6 and NGC~6357 when considered
together with the projected stellar kinematics and molecular-gas
evidence below.

\subsubsection{Plane-of-sky Kinematics and Formation Timescales}
\label{subsubsec:kin-timescale}

The approximately 1.8~Myr projected closest-approach epoch overlaps the
approximately $1$--$3$~Myr age range of the young clusters. Thus, the
epoch at which G352.8+0.6 approached the young-cluster region is
temporally consistent with the formation of the three young clusters,
providing a kinematic link between the older cluster and the recent
star-formation episode in NGC~6357. Because the traceback is restricted
to the plane of the sky, the three-dimensional configuration remains
unconstrained. We examined the available stellar radial velocities as
a possible independent check, but their large star-to-star scatter
and insufficient temporal information prevent reliable cluster
systemic velocities from being determined. We therefore judged them
unsuitable for validating the traceback; the measurements and their
reliability assessment are presented in Appendix~\ref{app:rv}.

\subsubsection{Possible Association with the \texorpdfstring{$+2$~\kms}{+2 km/s} Molecular Gas}
\label{subsubsec:plus2-association}

The $+2~\kms$ molecular component overlaps G352.8+0.6 in projection,
and its integrated intensity is locally reduced around the
spectroscopically identified massive stars belonging to G352.8+0.6. Stellar
radiation and winds can disperse molecular material
\citep{Krumholz2014,Dale2015}, so these features are consistent with
the $+2~\kms$ component having been affected by stars in the
G352.8+0.6 region. This spatial correspondence supports an association
between the molecular material and the stellar population.

Molecular material can survive in the larger-scale environment after gas
near a stellar cluster has been dispersed
\citep{Kawamura2009,Chevance2020}. We therefore interpret the
$+2~\kms$ component as a possible remnant of the parent molecular
material from which G352.8+0.6 formed.

The wide-field maps in Appendix~\ref{app:extended-co} place this
component in a broader molecular environment. Emission within the
$+2~\kms$ interval is prominent toward NGC~6334, and both the $-4$ and
$+2~\kms$ intervals contain emission across the region between the two
complexes, with complementary morphology in several regions. These
distributions suggest that the gas toward G352.8+0.6 may form part of
the larger-scale NGC~6334--NGC~6357 molecular environment.

Stellar kinematics have been used to infer the motions of
parental molecular gas in young star-forming regions
\citep{Getman2019,Zhou2025}.

\begin{figure*}[!tp]
\centering
\includegraphics[width=\textwidth]{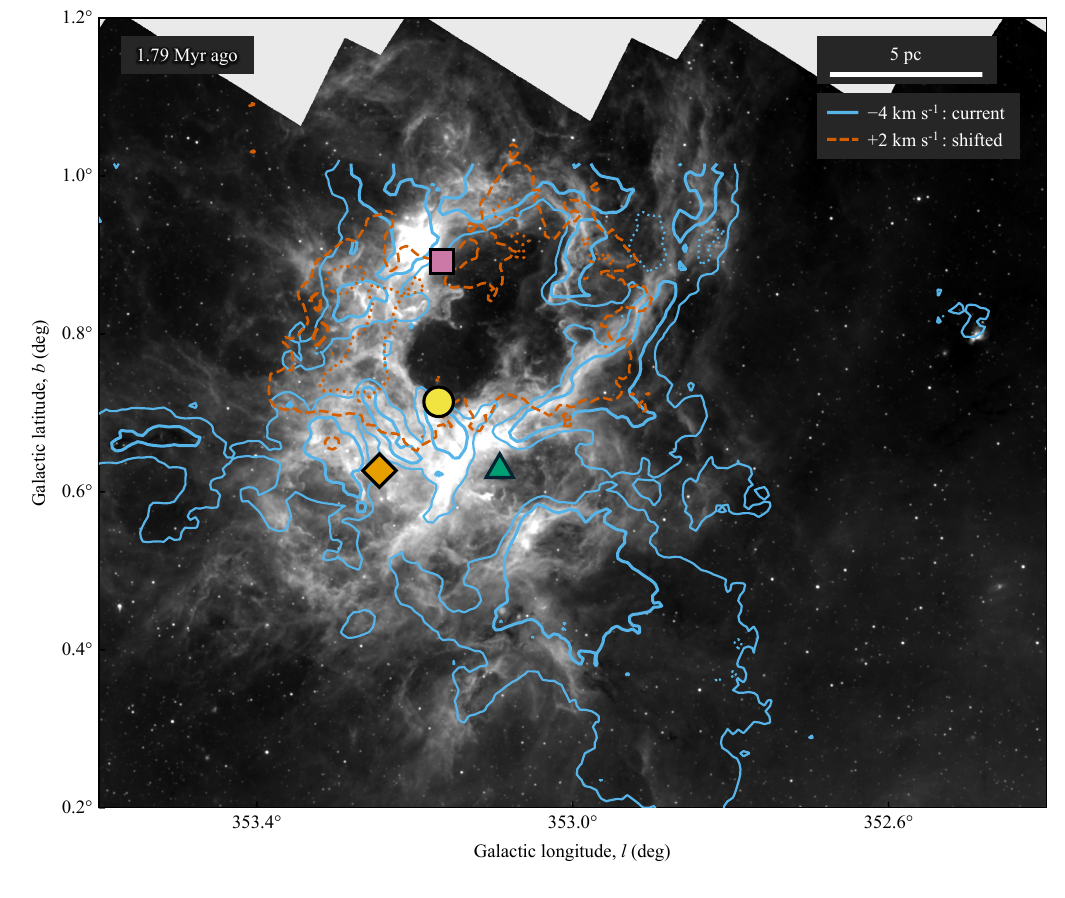}
\caption{
Projected reconstruction of the $+2~\kms$ molecular component
approximately 1.8~Myr ago (labeled 1.79~Myr in the figure), overlaid on
the present-day Spitzer/GLIMPSE $8~\micron$ image
\citep{Benjamin2003,Churchwell2009}. Blue solid contours show the current
$-4~\kms$ distribution, which is held fixed in the young-cluster
reference frame. Orange dashed contours show the present
$+2~\kms$ distribution translated backward using the relative proper
motion of G352.8+0.6, assuming that the cluster inherited and retained
the transverse motion of its parent gas. Contour levels and line
thicknesses follow Figure~\ref{fig:co-maps}; dotted segments enclose
local intensity depressions. The magenta square, green triangle,
and orange diamond mark the current centers of Pismis~24,
G353.1+0.6, and G353.2+0.6; the yellow circle marks the reconstructed
center of G352.8+0.6. The reconstruction translates the present gas
morphology at constant velocity. The white scale bar represents
approximately 5~pc.
}
\label{fig:gas-traceback}
\end{figure*}

In this interpretation, we assume that G352.8+0.6 inherited the
bulk plane-of-sky motion of its parent molecular gas and has
approximately retained that motion since its formation. We use the
present relative motion of G352.8+0.6 as a proxy for the past relative
transverse motion of the $+2~\kms$ component with respect to the
NGC~6357 gas. Displacing the present $+2~\kms$ distribution backward
by this motion over 1.8~Myr places it near Shell~A, where the present
$-4~\kms$ integrated intensity is relatively low
(Figure~\ref{fig:gas-traceback}). Under this assumption, the inferred
placement links the $+2~\kms$ material to the young NGC~6357 population
and Shell~A around the epoch indicated by the stellar traceback.

\subsection{Implications for the Formation of the Young Clusters and Shell~A}
\label{subsec:formation}

\subsubsection{A Molecular-cloud Collision Scenario}
\label{subsubsec:collision}

Molecular-cloud collisions can compress gas at an interaction interface
\citep{Habe1992,InoueFukui2013,Takahira2014}. Cloud penetration can
produce complementary projected distributions, while accelerated or
decelerated interface gas can produce intermediate-velocity emission
\citep{Takahira2014,Haworth2015,Fukui2021}. The complementary distributions of the
$-4$ and $+2~\kms$ components and the bridge-like intermediate-velocity
emission in the PV diagram are qualitatively consistent with these
interaction signatures (Figures~\ref{fig:co-maps} and \ref{fig:pv}).

\citet{Fukui2018} proposed a collision in NGC~6357 between components
near $-4$ and $-15~\kms$, based on complementary distributions and
bridge emission. Our scenario instead concerns the $-4$ and $+2~\kms$
structures and relates them to the independently measured stellar motion
of G352.8+0.6.

Combining these CO structures with the stellar traceback, we propose
a collision between the $+2$ and $-4~\kms$ molecular structures around
the characteristic traceback epoch, approximately 1.8~Myr ago. This
interpretation adopts the association of the $+2~\kms$ gas with
G352.8+0.6 and the stellar-motion proxy described in
Section~\ref{subsubsec:plus2-association}.

The 1.8~Myr closest-approach epoch represents a characteristic time
within an extended interaction that may have begun earlier. Such a
collision could have compressed the molecular gas and promoted the
formation of the young clusters and massive stars in NGC~6357, while
also contributing to the formation of Shell~A.

Following the proposed collision and star formation, ionizing
radiation and stellar winds from the massive stars, particularly those
in Pismis~24, may have reshaped Shell~A and modified the surrounding
gas motions. H$\alpha$ kinematic observations suggest that some
H~{\sc ii} regions in NGC~6357 have been shaped by stellar winds
\citep{Russeil2016}. In addition, \citet{Russeil2019} discussed expansion
of the central cavity, interpreting emission near $-9$ and
$+2~\kms$ as velocity extremes around a systemic velocity near
$-4~\kms$. This feedback-driven evolution can accompany the proposed
cloud collision and contribute to the present CO structure.

\begin{figure*}[!tp]
\centering
\includegraphics[width=\textwidth]{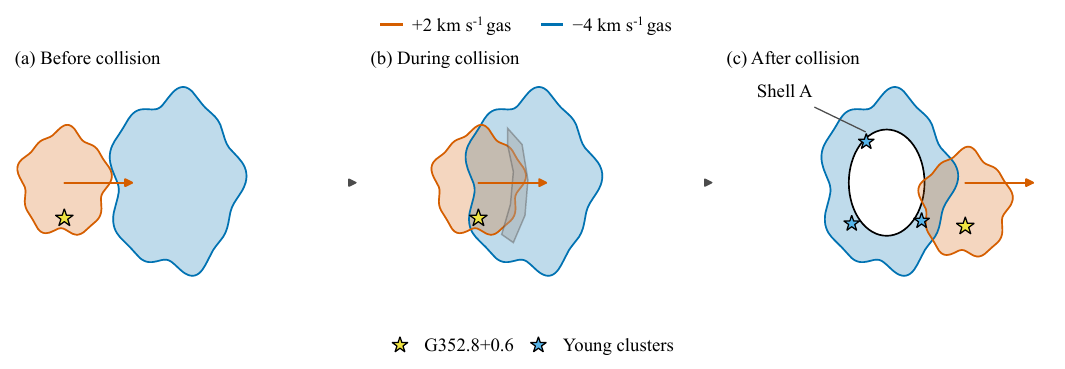}
\caption{
Schematic sequence of the proposed molecular-cloud collision
and subsequent evolution of NGC~6357. Orange and blue structures
represent the $+2$ and $-4~\kms$ molecular components, respectively.
The yellow star represents G352.8+0.6, and the blue stars represent
the three young clusters. (a) Before the collision, the $+2~\kms$
cloud and G352.8+0.6 approach the $-4~\kms$ cloud.
(b) During the interaction around the stellar traceback epoch,
approximately 1.8~Myr ago, gas is compressed at the gray interface,
promoting the formation of the young clusters and massive stars.
(c) After the interaction, the $+2~\kms$ cloud and the older cluster
move beyond the interaction region, leaving the shell-like cavity
marked by the black ellipse as Shell~A. Subsequent ionizing radiation
and stellar winds, particularly from Pismis~24, reshape the shell.
Orange arrows indicate the direction of relative cloud motion;
the small black arrowheads between panels indicate the sequence of
stages. Cloud shapes, sizes, and arrow lengths are schematic.
}
\label{fig:formation-scenario}
\end{figure*}

Figure~\ref{fig:formation-scenario} schematically summarizes this proposed evolutionary sequence, including the cloud collision, the formation of the young clusters, and the subsequent reshaping of Shell~A by stellar feedback.

\subsubsection{Alternative Scenarios}
\label{subsubsec:alternatives}

One alternative is prolonged star formation in the NGC~6357 region,
with G352.8+0.6 forming there approximately 8~Myr ago and subsequently
acquiring its present motion relative to the three young clusters.
In this scenario, G352.8+0.6 initially shared approximately the bulk
motion of the gas from which the younger clusters later formed.
Explaining the present relative transverse speed of approximately
$6.7~\kms$ requires a dynamical mechanism that changes the cluster's
bulk motion relative to that gas while maintaining the observed
proper-motion coherence of its members.

Another possibility is that G352.8+0.6, the $+2~\kms$ gas, and the young
NGC~6357 population are not all physically associated. This explanation
attributes the temporal correspondence between the stellar traceback
and young-cluster ages, the projected overlap of G352.8+0.6 with the
$+2~\kms$ gas, and the complementary and intermediate-velocity CO
structures to separate causes.

Among these possibilities, we retain molecular-cloud interaction as our
principal interpretation because it provides a unified explanation for
the stellar traceback, the young-cluster ages, and the complementary and
intermediate-velocity CO structures.

\subsection{Limitations and Future Tests}
\label{subsec:limitations}

Three principal limitations remain. First, residual differential parallax
systematics leave the relative line-of-sight positions uncertain, and the
lack of reliable systemic radial velocities restricts the traceback to
the plane of the sky. The approximately 1.8~Myr traceback therefore
constrains the epoch of closest projected approach rather than a
three-dimensional encounter.

Second, the gas displacement depends on the association of the
$+2~\kms$ component with the parent molecular material of G352.8+0.6
and on the cluster having inherited and approximately retained its
bulk plane-of-sky motion. The component's physical association and
transverse motion require independent constraints, and its subsequent
dynamical evolution is not included in the backward displacement.

Third, the optical depth of $^{12}$CO and line-of-sight overlap affect
the interpretation of the molecular structure \citep{Bolatto2013}.
Variations in column density, excitation, and optical depth also affect
CO intensity \citep{MangumShirley2015}. The formation scenario therefore
uses the spatial and velocity structures together with the stellar
kinematics.

Multi-epoch high-resolution spectroscopy of cluster members can constrain
their systemic radial velocities while reducing the effects of unresolved
binaries, enabling three-dimensional stellar kinematics and a direct
comparison with the molecular-gas radial velocities. Higher angular- and
spectral-resolution observations of CO isotopologues such as
$^{13}$CO and C$^{18}$O can better resolve the molecular structure and
separate overlapping velocity components. Finally, numerical simulations
tailored to the observed geometry can test whether interaction between
the $-4$ and $+2~\kms$ molecular structures can reproduce the present
gas distribution, the formation timescale of the young clusters, and
Shell~A.

\section{Summary and Conclusions}
\label{sec:summary}

We investigated the past star-formation process in NGC~6357 by combining
Gaia DR3 astrometry with ThrUMMS molecular-gas data. Our main results are:

\begin{enumerate}

\item
The Gaia HDBSCAN search independently recovered the cataloged cluster
G352.8+0.6 alongside Pismis~24, G353.1+0.6, and G353.2+0.6.
G352.8+0.6 is approximately 8~Myr old, compared with approximately
1--3~Myr for the three young clusters. The inferred distances of all four
clusters lie near the mean heliocentric distance of the G352 complex,
$1.67\pm0.08$~kpc \citep{Povich2026}.
G352.8+0.6 has the smallest nominal distance, approximately 0.05--0.13~kpc closer than
the three young clusters.

\item
G352.8+0.6 has a distinct plane-of-sky motion relative to the three
young clusters. A linear traceback places its closest projected approach
to the young-cluster reference center approximately 1.8~Myr ago,
overlapping the inferred formation ages of the young clusters.

\item
The $+2~\kms$ CO component overlaps G352.8+0.6 and shows localized
intensity depressions around the three spectroscopically identified
massive stars belonging to G352.8+0.6. The $-4$ and $+2~\kms$ components show complementary
projected distributions, and CO emission extends through intermediate
velocities between the two components in the PV diagram.

\item
We interpret the $+2~\kms$ component as a possible remnant of the
molecular material from which G352.8+0.6 formed. Assuming that the
cluster inherited and approximately retained the bulk plane-of-sky
motion of this parent gas, its present relative motion can be used as a
proxy for the past relative transverse motion of the $+2~\kms$
component with respect to the NGC~6357 gas. Under this interpretation,
we propose that the $+2$ and $-4~\kms$ molecular structures interacted
around the epoch indicated by the stellar traceback, approximately
1.8~Myr ago. The interaction may have begun earlier than this nominal
closest-approach epoch and may have promoted the formation of the young
clusters and massive stars and contributed to the formation of Shell~A.
Subsequent ionizing radiation and stellar winds may then have reshaped
the shell.

\end{enumerate}

We adopt this molecular-cloud interaction scenario as the principal
interpretation of the stellar and molecular-gas observations. Its key
remaining uncertainties are the three-dimensional association of
G352.8+0.6 with the young NGC~6357 population and the physical
association and transverse motion of the $+2~\kms$ molecular gas.
Reliable systemic stellar radial velocities and further molecular-line
observations will provide direct tests of these points.

More broadly, this study demonstrates how the kinematics of an older
stellar population can provide a complementary probe of past
star-formation processes that are difficult to reconstruct from the
youngest stars alone.

\begin{acknowledgments}

This work made use of data from the European Space Agency (ESA) mission
\textit{Gaia}, processed by the \textit{Gaia} Data Processing and Analysis
Consortium (DPAC). Funding for DPAC is provided by national institutions,
in particular those participating in the \textit{Gaia} Multilateral
Agreement.

This work made use of the VIRAC2 source catalogue, based on data obtained
from the ESO Science Archive Facility
(DOI: 10.18727/archive/68).

This work is based in part on observations made with the
\textit{Spitzer Space Telescope}, operated by the Jet Propulsion Laboratory,
California Institute of Technology, under contract with NASA.
This research also made use of the NASA/IPAC Infrared Science Archive,
operated by the California Institute of Technology under contract with NASA.

The H$\alpha$ imaging used in this work is based on data products from the
VST Photometric H$\alpha$ Survey of the Southern Galactic Plane and Bulge
(VPHAS+), obtained with ESO telescopes under public survey programme
177.D-3023.

We gratefully acknowledge the support provided by the
Super Science High School (SSH) program.

\end{acknowledgments}

\section*{AI/LLM Disclosure}

OpenAI's ChatGPT\footnote{\url{https://chatgpt.com/}} was used under author supervision to assist with Python code
drafting and revision, literature-search support and translation,
and English-language editing and organization of the manuscript. AI-assisted
outputs used in the research or manuscript were checked by the authors
against the underlying code, data, and original literature. Responsibility
for the scientific analysis, interpretation, and final manuscript lies
entirely with the authors.

\facilities{
Gaia,
Mopra,
Spitzer,
VST (OmegaCAM),
ESO:VISTA,
VLT:Antu (KMOS)
}

\software{
hdbscan \citep{McInnes2017},
gaiadr3-zeropoint \citep{Lindegren2021},
dustapprox (version 0.1) \citep{Fouesneau2022},
Astropy \citep{Astropy2022},
NumPy \citep{Harris2020},
SciPy \citep{Virtanen2020},
pandas \citep{McKinney2010},
Matplotlib \citep{Hunter2007},
scikit-learn \citep{Pedregosa2011},
astroquery \citep{Ginsburg2019}
}

\begin{contribution}
G.T.W. led the data analyses, interpreted the results,
and wrote the manuscript.
S.N. provided scientific guidance, assisted with the analyses,
contributed to the interpretation of the results,
and revised the manuscript.
K.T. provided feedback on the research and reviewed
and revised the manuscript.
T.O. supervised the research and provided advice based on progress reports from the first author. 
\end{contribution}

\appendix
\numberwithin{figure}{section}
\renewcommand{\thefigure}{\thesection\arabic{figure}}
\renewcommand{\theHfigure}{\thesection.\arabic{figure}}
\renewcommand{\theHequation}{appendix.\thesection.\arabic{equation}}

\section{Additional Cluster Candidates and Target Selection}
\label{app:candidates}

The five-dimensional Gaia DR3 HDBSCAN analysis identified 3733 candidate
overdensities over the full $3^\circ$ search field.
For the NGC~6357 follow-up region, we selected candidates having at
least one member within
\begin{equation}
352.6^\circ\leq l\leq353.6^\circ,
\qquad
0.2^\circ\leq b\leq1.2^\circ.
\end{equation}
This criterion uses member positions, so a candidate center may lie
outside the rectangle. No parallax or distance restriction was applied
to the candidate listing in this appendix.

\begin{figure*}[!tp]
\centering
\includegraphics[width=\textwidth]{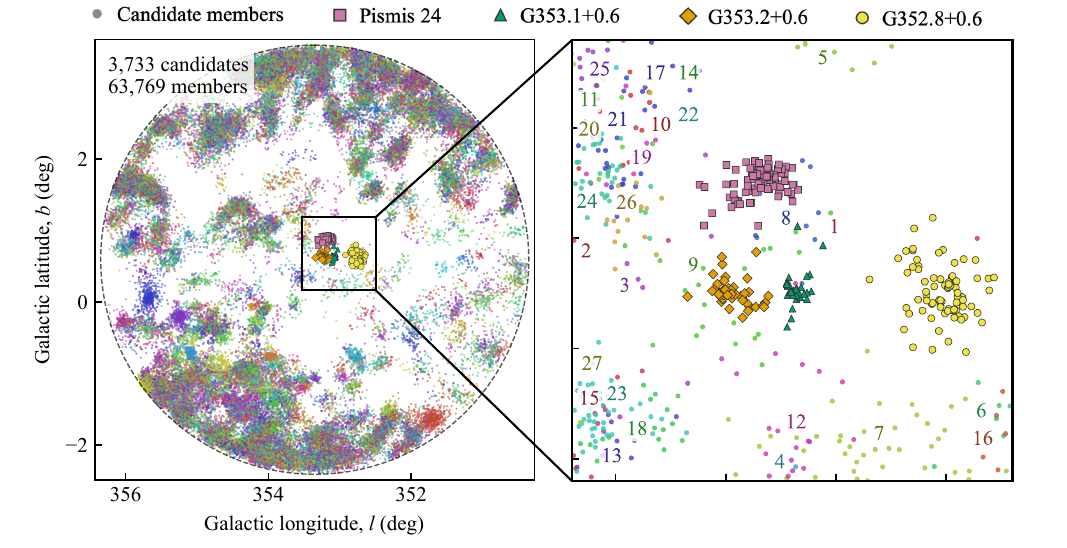}
\caption{
Spatial distribution of the Gaia DR3 HDBSCAN candidates in
Galactic coordinates. Left: the full $3^\circ$ search field, containing
3733 candidate overdensities and 63,769 assigned members. Small dots
show individual candidate members, with colors distinguishing candidate
groups; the dashed boundary outlines the search field.
The black rectangle and connecting lines identify the enlarged region
on the right, spanning $352.6^\circ\leq l\leq353.6^\circ$ and
$0.2^\circ\leq b\leq1.2^\circ$.
Magenta squares, green triangles, orange diamonds, and yellow circles
highlight members of Pismis~24, G353.1+0.6, G353.2+0.6, and G352.8+0.6,
respectively, in both panels. The 27 additional candidates in the
enlarged region are numbered as in Table~\ref{tab:app-candidates};
the numbers are display identifiers, not HDBSCAN label IDs.
Candidates enter this listing if at least one member lies within the
rectangle, without a parallax or distance restriction.
}
\label{fig:app-candidates-full}
\label{fig:app-candidates-ngc6357}
\end{figure*}

The spatial criterion selects 31 candidates. Four---Pismis~24,
G353.1+0.6, G353.2+0.6, and G352.8+0.6---are the principal targets
analyzed in the main text and listed in
Table~\ref{tab:cluster-properties}. The remaining 27 candidates are
shown in Figure~\ref{fig:app-candidates-full} and listed in
Table~\ref{tab:app-candidates}.

\begin{deluxetable*}{lccc}
\tabletypesize{\footnotesize}
\tablecaption{
Additional Gaia HDBSCAN Candidates in the NGC~6357 Follow-up Region
\label{tab:app-candidates}
}
\tablehead{
\colhead{Candidate} & \colhead{Distance} & \colhead{Age} & \colhead{$A_V$} \\
\colhead{} & \colhead{(kpc)} & \colhead{(Myr)} & \colhead{(mag)}
}
\startdata
1 & $0.13^{+0}_{-0}$ & $31.4^{+13.5}_{-9.3}$ & $0.89^{+0.50}_{-0.29}$ \\
2 & $1.06^{+0.18}_{-0.14}$ & $50.0^{+23.4}_{-21.0}$ & $1.26^{+0.32}_{-0.43}$ \\
3 & $1.03^{+0.03}_{-0.03}$ & $52.9^{+20.8}_{-24.6}$ & $1.43^{+0.32}_{-0.50}$ \\
4 & $1.16^{+0.04}_{-0.04}$ & $46.2^{+29.2}_{-24.0}$ & $2.30^{+0.67}_{-0.81}$ \\
5 & $0.99^{+0.02}_{-0.02}$ & $36.0^{+20.6}_{-14.1}$ & $1.82^{+0.18}_{-0.21}$ \\
6 & $1.31^{+0.13}_{-0.11}$ & $14.0^{+37.8}_{-9.9}$ & $3.88^{+0.80}_{-1.17}$ \\
7 & $1.05^{+0.01}_{-0.01}$ & $33.8^{+7.7}_{-6.4}$ & $2.11^{+0.08}_{-0.08}$ \\
8 & Negative parallax\tablenotemark{a} & \nodata & \nodata \\
9 & $1.04^{+0.02}_{-0.02}$ & $28.9^{+24.8}_{-12.9}$ & $2.38^{+0.44}_{-0.36}$ \\
10 & $1.23^{+0.05}_{-0.05}$ & $62.0^{+23.1}_{-18.1}$ & $2.27^{+0.51}_{-0.53}$ \\
11 & $1.88^{+0.36}_{-0.26}$ & $9.0^{+29.5}_{-7.0}$ & $3.63^{+1.82}_{-1.93}$ \\
12 & $1.01^{+0.01}_{-0.01}$ & $30.8^{+17.0}_{-14.5}$ & $1.93^{+0.33}_{-0.44}$ \\
13 & $0.88^{+0.03}_{-0.03}$ & $60.5^{+24.7}_{-21.7}$ & $1.61^{+0.90}_{-0.74}$ \\
14 & $1.56^{+0.08}_{-0.07}$ & $2.6^{+11.7}_{-1.4}$ & $2.01^{+2.20}_{-0.86}$ \\
15 & $1.76^{+0.22}_{-0.18}$ & $12.0^{+31.4}_{-8.3}$ & $3.96^{+0.94}_{-1.18}$ \\
16 & $1.63^{+0.05}_{-0.05}$ & $4.7^{+5.6}_{-2.6}$ & $5.26^{+0.56}_{-0.92}$ \\
17 & $1.54^{+0.07}_{-0.07}$ & $21.9^{+39.3}_{-17.3}$ & $3.78^{+0.96}_{-1.13}$ \\
18 & $1.32^{+0.11}_{-0.09}$ & $9.0^{+33.6}_{-6.9}$ & $3.94^{+1.60}_{-2.39}$ \\
19 & $9.57^{+24.87}_{-4.01}$ & $3.7^{+23.2}_{-1.8}$ & $7.27^{+0.40}_{-0.64}$ \\
20 & $3.23^{+0.59}_{-0.43}$ & $9.7^{+37.0}_{-6.8}$ & $8.11^{+0.54}_{-0.33}$ \\
21 & $4.08^{+0.67}_{-0.51}$ & $4.1^{+53.9}_{-2.0}$ & $7.09^{+0.29}_{-0.41}$ \\
22 & $3.41^{+0.94}_{-0.61}$ & $3.0^{+23.6}_{-1.7}$ & $7.49^{+0.26}_{-0.37}$ \\
23 & $1.80^{+0.09}_{-0.08}$ & $2.0^{+1.2}_{-0.6}$ & $5.35^{+0.34}_{-0.31}$ \\
24 & $3.03^{+0.34}_{-0.28}$ & $21.6^{+29.8}_{-13.4}$ & $6.84^{+0.15}_{-0.27}$ \\
25 & $1.64^{+0.04}_{-0.04}$ & $3.6^{+4.4}_{-2.4}$ & $2.73^{+1.13}_{-1.01}$ \\
26 & $1.65^{+0.07}_{-0.07}$ & $5.9^{+9.5}_{-4.1}$ & $3.93^{+0.77}_{-1.81}$ \\
27 & $2.53^{+0.33}_{-0.26}$ & $17.2^{+35.0}_{-12.7}$ & $7.19^{+0.69}_{-0.74}$
\enddata
\tablenotetext{a}{Candidate 8 has a negative parallax. No distance, age,
or extinction estimate is reported.}
\tablecomments{
Numbers identify the 27 additional Gaia HDBSCAN candidates in
the right panel of Figure~\ref{fig:app-candidates-full}; they are not
HDBSCAN label IDs. A candidate is included if at least one member lies
within $352.6^\circ\leq l\leq353.6^\circ$ and
$0.2^\circ\leq b\leq1.2^\circ$, without a parallax or distance cut.
The four principal clusters are listed separately in
Table~\ref{tab:cluster-properties}. $A_V$ retains the approximation
$A_V\simeq A_0$ used throughout the manuscript. Zero entries in the
reported distance uncertainties reflect rounding.
}
\end{deluxetable*}

\section{Gaia DR3--VIRAC2 Comparison and the Non-Recovery of G352.8+0.6 in VIRAC2}
\label{app:virac2}

\subsection{Gaia--VIRAC2 Cross-matching}
\label{app:virac2-crossmatch}

We cross-matched the 257 Gaia DR3 members of the four principal clusters
with VIRAC2 \citep{Smith2025}, using VizieR catalog
\texttt{II/387/virac2}. Gaia positions were linearly propagated from
J2016.0 to the VIRAC2 reference epoch J2014.0 using the Gaia proper
motions. We used CDS XMatch to retrieve all candidate counterparts
within an angular separation of $0.7\arcsec$ at this common epoch.

In addition to positional agreement, we evaluated the difference between
the catalog proper motions using
\begin{equation}
\begin{aligned}
D_\mu = \Biggl[&
\frac{(\mu_{\alpha*,G}-\mu_{\alpha*,V})^2}
{\sigma_{\alpha*,G}^{2}+\sigma_{\alpha*,V}^{2}} \\
&+\frac{(\mu_{\delta,G}-\mu_{\delta,V})^2}
{\sigma_{\delta,G}^{2}+\sigma_{\delta,V}^{2}}
\Biggr]^{1/2},
\end{aligned}
\label{eq:virac2-pm-match}
\end{equation}
where the subscripts $G$ and $V$ denote Gaia and VIRAC2,
$\mu_{\alpha*}=\dot{\alpha}\cos\delta$, and the $\sigma$ terms are the
reported uncertainties of the corresponding proper-motion components.
The measurement variances are summed component by component; covariance
between the two proper-motion components is not included.
Candidates were accepted if their angular separation was no greater
than $0.7\arcsec$ and $D_\mu\leq5$.

If more than one candidate passed these criteria for a Gaia source,
we retained the candidate with the smallest angular separation.
If multiple Gaia sources were associated with the same VIRAC2 source,
we likewise retained the association with the smallest angular
separation. This procedure yielded one-to-one VIRAC2 counterparts for
202 of the 257 Gaia members.

\subsection{Comparison of the Gaia DR3 and VIRAC2 Cluster Searches}
\label{app:virac2-comparison}

To compare the optical and near-infrared astrometric cluster searches using the same set of observables, we performed four-dimensional HDBSCAN analyses in $(\alpha,\delta,\mu_{\alpha *},\mu_{\delta})$ for both Gaia DR3 and VIRAC2. Each variable was standardized independently within the corresponding input catalog. For both analyses, we adopted \texttt{min\_cluster\_size}=8, \texttt{min\_samples}=8, Euclidean distance, and the leaf cluster-selection method, as described in Section~\ref{subsec:hdbscan}.

The four-dimensional Gaia DR3 analysis recovered a group corresponding to G352.8+0.6 even when parallax was excluded from the clustering space. In contrast, the VIRAC2 analysis recovered candidate groups corresponding to the three young clusters, Pismis~24, G353.1+0.6, and G353.2+0.6, but did not identify G352.8+0.6 as a distinct cluster candidate. The difference therefore persists when the comparison is restricted to the four astrometric quantities available with useful precision in both catalogs.

Of the 76 Gaia-selected members of G352.8+0.6, 73 have VIRAC2
counterparts under the criteria in
Appendix~\ref{app:virac2-crossmatch}, and all 73 have finite
proper-motion and epoch information. Thus, nearly all of the
Gaia-selected population is represented in VIRAC2.

\begin{figure*}[!tp]
\centering
\includegraphics[width=\textwidth,height=0.75\textheight,keepaspectratio]{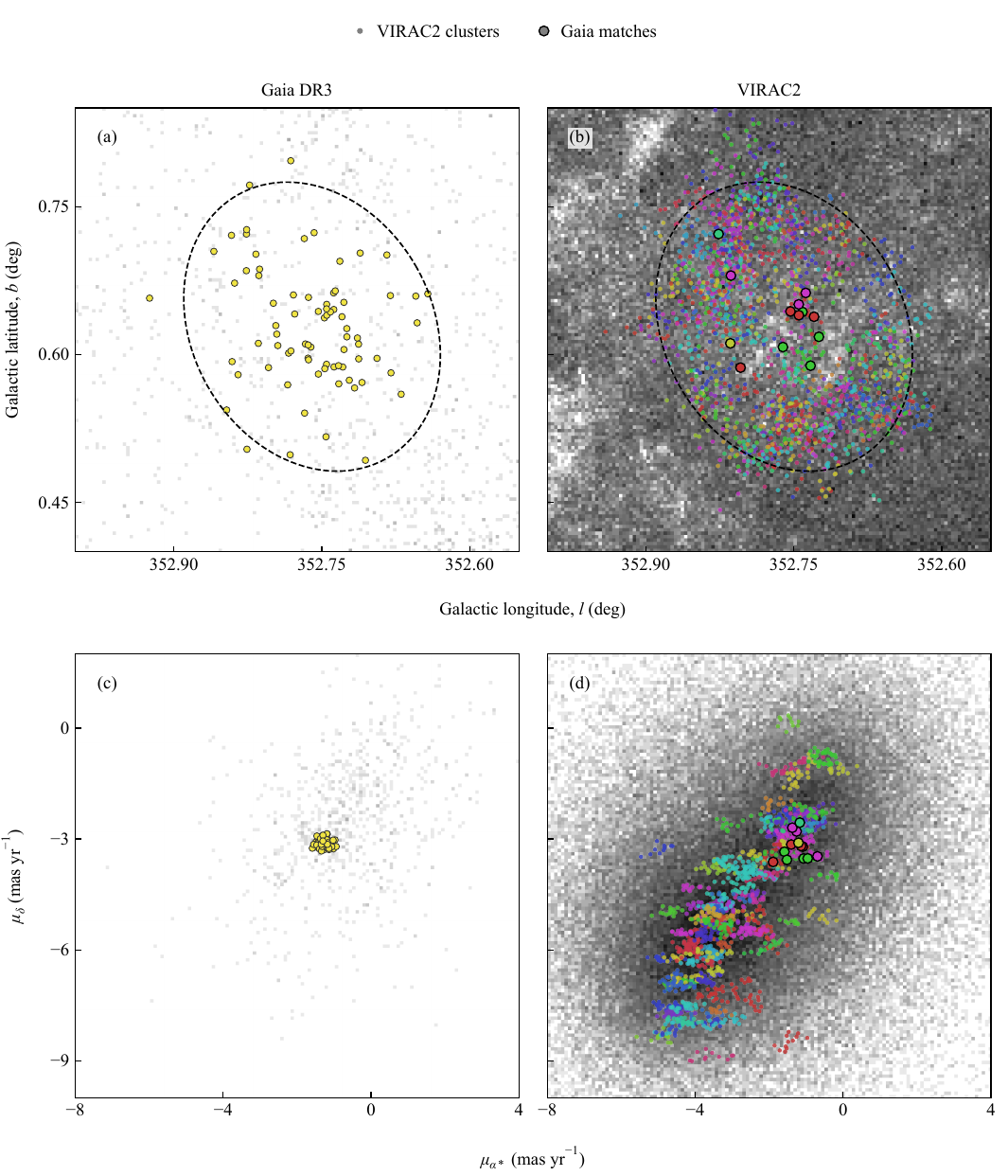}
\caption{
Spatial and proper-motion comparison of Gaia DR3 and VIRAC2
around G352.8+0.6. The left column shows Gaia DR3 and the right
column shows VIRAC2. The upper panels, (a) and (b), show positions
in Galactic coordinates $(l,b)$; the lower panels, (c) and (d),
show catalog proper motions $(\mu_{\alpha*},\mu_\delta)$ in
mas~yr$^{-1}$, where $\mu_{\alpha*}=\mu_\alpha\cos\delta$.
Both rows use the sky region shown in the upper panels.
In (a) and (c), yellow circles mark the adopted Gaia members of
G352.8+0.6. In (b) and (d), small colored points show members of
VIRAC2 HDBSCAN candidate groups whose mean sky positions lie
within the dashed ellipse in (b); colors distinguish groups and
are consistent between the two panels. Larger circles with black
outlines highlight the counterparts of the adopted Gaia members
within these displayed VIRAC2 groups, retaining their group colors.
The grayscale backgrounds show binned counts of catalog sources
after excluding the adopted Gaia members and their VIRAC2
counterparts. Darker pixels indicate higher counts, with the same
logarithmic display scale for the two catalogs within each row.
The identical dashed ellipses in (a) and (b) enclose approximately
95\% of the adopted Gaia population and define the aperture used
for the source-density comparison.
}
\label{fig:app-virac2}
\end{figure*}

Figure~\ref{fig:app-virac2} compares the spatial and proper-motion
distributions in Gaia DR3 and VIRAC2 around G352.8+0.6.

\subsection{Source Density and Cluster Contrast}
\label{app:virac2-density}

The Gaia-selected stellar populations differ substantially in their projected surface number densities. We define
\begin{equation}
\Sigma_{90}=\frac{0.9N}{\pi r_{90}^{2}},
\end{equation}
where $N$ is the number of selected Gaia members and $r_{90}$ is the projected radius enclosing 90\% of them. Using the values in Table~\ref{tab:cluster-properties}, we obtain approximately 4.7, 2.8, 2.4, and 1.8 stars pc$^{-2}$ for Pismis~24, G353.1+0.6, G353.2+0.6, and G352.8+0.6, respectively. G352.8+0.6 therefore has the lowest mean projected member density among the four principal clusters in the Gaia-selected sample.

We use $\Sigma_{90}$ as an empirical measure of the relative compactness
of the Gaia-selected populations. The member samples reflect the clusters'
distances, extinctions, ages, catalog completeness, and member selection.

The cataloged source density around G352.8+0.6 is substantially higher
in VIRAC2 than in Gaia DR3. Within the same elliptical aperture, the
VIRAC2 source count is approximately 30 times the Gaia DR3 source count.
The counts include all catalog sources in the aperture, and their ratio
measures the difference in projected catalog source density.

In this Galactic-plane field, the lower extinction at near-infrared
wavelengths allows VIRAC2 to sample more heavily reddened stars,
including background populations suppressed in Gaia's optical sample.
This can increase the unrelated field population and reduce the
contrast of a diffuse cluster.

The relatively diffuse Gaia-selected cluster population and the much
denser VIRAC2 field provide a plausible explanation for the lower
clustering contrast of G352.8+0.6 in VIRAC2. This is a qualitative
interpretation of the projected-density comparison. Proper-motion
uncertainties and catalog-dependent feature scaling can also affect
the recovery in four-dimensional astrometric space.

\subsection{Relation to Previous Surveys of NGC 6357}
\label{app:virac2-previous}

G352.8+0.6 has an inferred age of approximately 8 Myr, compared with
approximately 1--3 Myr for the three principal young clusters. As clusters
evolve and their natal material is dispersed, their stellar populations
can become more spatially extended \citep{LadaLada2003}. Its relatively
large $r_{90}$ and low projected member density are qualitatively
consistent with a more evolved stellar population.

Previous studies of NGC 6357 have often emphasized very young stellar populations identified through diagnostics such as infrared excess and X-ray emission. The incidence of infrared excess decreases as circumstellar disks dissipate \citep{Haisch2001}, while stellar X-ray properties also evolve with age and stellar mass \citep{Preibisch2005}. An older population such as G352.8+0.6 may therefore be less conspicuous in samples optimized for identifying the youngest stars and embedded populations.

G352.8+0.6 was already cataloged as PHOC~17 and UFMG~76
\citep{HuntReffert2021,Ferreira2021}. Our analysis examines its physical
and kinematic relationship with the younger NGC~6357 population.
Previous studies have primarily focused on young stellar populations
and active star-forming structures \citep{Fang2012,Russeil2020}.
The older age and relatively extended distribution of G352.8+0.6 may
help explain why its connection to NGC~6357 has received comparatively
little attention.

\section{Parallax Systematics and Relative Cluster Distances}
\label{app:parallax}

The nominal Gaia parallaxes place G352.8+0.6 approximately 50--130~pc
in front of the three principal young clusters
(Section~\ref{subsubsec:distance}). We examine the sensitivity of this
relative placement directly in parallax space, where residual Gaia
astrometric systematics are expressed, before converting to distance
through $d=1/\varpi$.

Restricting the member samples to the published calibration ranges
of the parallax zero-point correction
(Section~\ref{subsec:data-gaia}) changes the nominal cluster distances
by approximately $-16$, $+3$, $-4$, and $-16$~pc for Pismis~24,
G353.1+0.6, G353.2+0.6, and G352.8+0.6, respectively.
This restricted-sample comparison provides a sensitivity test of
the nominal distances.

For each young cluster $j$, we define the difference between the median
zero-point-corrected parallaxes as
\begin{equation}
\Delta\varpi_j
=
\tilde{\varpi}_{\rm G352.8+0.6}
-
\tilde{\varpi}_{j},
\label{eq:delta-parallax}
\end{equation}
where $\tilde{\varpi}$ denotes the median corrected parallax of the
selected members. A positive $\Delta\varpi_j$ therefore indicates that
G352.8+0.6 has the larger nominal parallax and hence the smaller nominal
heliocentric distance.

Using the representative parallaxes adopted in the present analysis,
\begin{equation}
\tilde{\varpi}_{\rm G352.8+0.6}=0.632414~{\rm mas},
\end{equation}
while the corresponding values for Pismis~24, G353.1+0.6, and
G353.2+0.6 are $0.594145$, $0.585796$, and $0.612282$~mas,
respectively.
The nominal pairwise differences are therefore
\begin{equation}
\Delta\varpi_j
=
0.038269,\ 0.046617,\ {\rm and}\ 0.020132~{\rm mas},
\end{equation}
for Pismis~24, G353.1+0.6, and G353.2+0.6, respectively.

We constructed each pairwise distribution from the corresponding
independently resampled member lists using the same 20,000-realization
bootstrap prescription as the main distance analysis. The
16th--84th percentile intervals on $\Delta\varpi_j$ are
$[0.02693,0.04673]$, $[0.02942,0.05397]$, and
$[0.00899,0.03744]$~mas for Pismis~24, G353.1+0.6, and
G353.2+0.6, respectively. The corresponding 2.5th--97.5th percentile
intervals are $[0.01950,0.05418]$, $[0.01804,0.05801]$, and
$[-0.00010,0.05036]$~mas. Thus, the last pair's 95\% bootstrap
interval already includes zero without an imposed systematic shift.

To examine the sensitivity to differential residual systematics, we
apply opposing test shifts to the representative parallaxes in each pair.
For a trial shift $\delta\varpi$,
\begin{equation}
\tilde{\varpi}_{j}
\rightarrow
\tilde{\varpi}_{j}
+
\delta\varpi
\end{equation}
and
\begin{equation}
\tilde{\varpi}_{\rm G352.8+0.6}
\rightarrow
\tilde{\varpi}_{\rm G352.8+0.6}
-
\delta\varpi.
\end{equation}
The relative parallax difference then becomes
\begin{equation}
\Delta\varpi_j
\rightarrow
\Delta\varpi_j
-
2\delta\varpi.
\label{eq:parallax-test-shift}
\end{equation}

Thus, an opposing shift of $0.02$~mas per cluster changes the pairwise
parallax difference by $0.04$~mas. Here $\delta\varpi$ is the per-cluster
test shift and $2\delta\varpi$ is the pairwise change. The imposed shifts
are sensitivity-test parameters rather than estimates of actual
differential Gaia residual offsets.

For reference, the opposing per-cluster shifts required to make the
\emph{nominal} representative parallaxes equal are
\begin{equation}
\delta\varpi_{\rm equal}
=
\frac{\Delta\varpi_j}{2},
\end{equation}
which gives approximately $0.01913$, $0.02331$, and $0.01007$~mas for
Pismis~24, G353.1+0.6, and G353.2+0.6, respectively.

With shifts scanned in $0.0001$~mas steps, the 16th--84th percentile
interval first includes zero at an opposing per-cluster shift of
$0.0135$, $0.0148$, and $0.0045$~mas for Pismis~24, G353.1+0.6,
and G353.2+0.6, respectively.

The corresponding heliocentric distances can be recalculated from the
shifted representative parallaxes using the same estimator as in the
principal analysis.

\begin{figure}[!tp]
\centering
\includegraphics[width=\columnwidth]{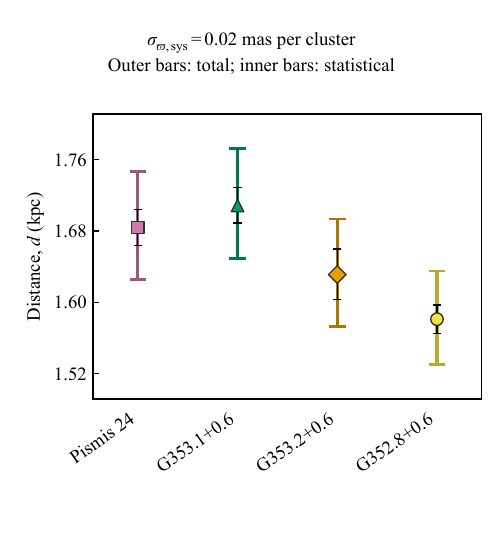}
\caption{
Cluster distances and an illustrative parallax-error budget.
Magenta squares, green triangles, orange diamonds, and yellow circles
represent Pismis~24, G353.1+0.6, G353.2+0.6, and G352.8+0.6,
respectively. The distances are reciprocals of the median
zero-point-corrected parallaxes of the quality-screened members
described in this appendix. Thin black inner bars propagate the
MAD-based statistical uncertainty of each cluster median. Thick colored
outer bars additionally include a cluster-level systematic term
$\sigma_{\varpi,\mathrm{sys}}=0.02$~mas, added in quadrature in
parallax space before conversion to distance. The systematic term is
applied to each cluster as a whole. These illustrative error bars use
a different prescription from the member-bootstrap intervals in
Table~\ref{tab:cluster-properties}.
}
\label{fig:app-parallax}
\end{figure}

Figure~\ref{fig:app-parallax} provides a complementary illustration of
the cluster distances with a cluster-level parallax-error budget.
For this figure, we use members with RUWE $\leq1.4$, five- or
six-parameter astrometric solutions, and positive corrected parallaxes
and parallax uncertainties. The retained counts are 100, 32, 46, and
76 for Pismis~24, G353.1+0.6, G353.2+0.6, and G352.8+0.6,
respectively. The representative parallax is the median of each
quality-screened sample. Its statistical uncertainty is estimated as
\begin{equation}
\sigma_{\varpi,\mathrm{stat}}
=\frac{1.253\times1.4826\,
\mathrm{MAD}(\varpi_{\rm corr})}{\sqrt{N}},
\end{equation}
where MAD is the median absolute deviation from the sample median.
For the outer bars, we adopt $\sigma_{\varpi,\mathrm{sys}}=0.02$~mas
per cluster and combine it with the statistical term in quadrature.
The endpoints are converted to distance through
$[1/(\tilde{\varpi}+\sigma_\varpi),
1/(\tilde{\varpi}-\sigma_\varpi)]$ in kpc when parallaxes are in mas.
The inner bars use $\sigma_{\varpi,\mathrm{stat}}$ alone.
This illustrative error budget is separate from the opposing-shift
test above and the full-member bootstrap intervals reported in
Table~\ref{tab:cluster-properties}.

Only the differential residual systematic between a pair of clusters
affects their relative parallax. If the residual systematic offsets are
$s_{\rm G352.8+0.6}$ and $s_j$, respectively, their contribution to
Equation~(\ref{eq:delta-parallax}) is
\begin{equation}
s_{\rm G352.8+0.6}-s_j.
\end{equation}
A systematic component common to both clusters therefore cancels in the
relative parallax. Because the clusters occupy nearby lines of sight,
spatially correlated Gaia systematics may produce such common-mode
errors. Conversely, differences in position, magnitude, color, and
astrometric properties of the member samples can leave differential
residual offsets even within the same general region.

The inferred relative line-of-sight separations change substantially
under differential parallax offsets of the tested magnitude. This
sensitivity leaves a physical association with NGC~6357 viable when the
distance estimates are considered alongside the stellar kinematics
and molecular-gas structure.

\section{Stellar Radial-Velocity Measurements}
\label{app:rv}

For the KMOS cross-match, we used the radial velocities and reported
uncertainties of the 22 NGC~6357 stars in Table B.2 of
\citet{RamirezTannus2021}. We joined this table by the literature target
names to the NGC~6357 target catalog of \citet{RamirezTannus2020}.
Associations with Gaia members required exact agreement of the
near-infrared source identifiers linked to the Gaia member catalog,
with each identifier unique within the full HDBSCAN member sample.

For positional verification, we assumed a comparison epoch of J2000.0
and linearly propagated the Gaia DR3 positions from J2016.0 using their
proper motions. An identifier-based association was accepted only if
it agreed with the positional nearest-neighbor association, the two
sources were mutually nearest neighbors, and their angular separation
was no greater than $1.0\arcsec$. We also required a Gaia magnitude
$G\leq19$~mag and an absolute difference of no more than 0.05~mag
between the literature target's $K_s$ magnitude and the near-infrared
photometry linked to its Gaia counterpart.

After acceptance, we verified that the absolute magnitude differences
in each of $J$, $H$, and $K_s$ were no greater than 0.0051~mag.
Fifteen of the 22 KMOS sources were accepted; all 15 belong to the
four principal clusters. The largest angular separation among the
accepted associations was $0.281\arcsec$.

We compiled available radial-velocity information for Gaia-selected members from Gaia DR3 and published VLT/KMOS spectroscopy described in Section~\ref{subsec:data-ancillary}, with the aim of assessing whether reliable cluster systemic velocities could be derived from the existing measurements. The compilation contains 38 catalog records for 37 unique stars: 15 in Pismis~24, 6 in G353.1+0.6, 13 in G353.2+0.6, and 3 in G352.8+0.6. It comprises 23 Gaia catalog aggregates and 15 KMOS photospheric measurements, with one star appearing in both sources. Records and unique stars are distinguished explicitly, while catalog-reported repeat counts are treated separately from independently verified observing epochs.

Table~\ref{tab:app-rv} lists the reported values, uncertainties, catalog sources, reference-frame information, and available warning information. Gaia DR3 velocities are catalog-level aggregates rather than individual single-epoch measurements. Reported transit or visit counts therefore do not establish how many independent dated RV measurements are available for a variability assessment. The available records do not provide a verified dated repeated-RV series for these stars, so the repeat counts alone cannot be used to determine whether a catalog velocity represents the systemic velocity of an individual star.

\begin{deluxetable*}{llrrllrl}
\tabletypesize{\scriptsize}
\tablecaption{
38 Radial-velocity Records for 37 Unique Gaia-selected Cluster Members
\label{tab:app-rv}
}
\tablehead{
\colhead{Gaia source ID} & \colhead{Cluster} & \colhead{RV} &
\colhead{$\sigma_{\rm RV}$} & \colhead{Source} & \colhead{Frame} &
\colhead{$n_{\rm rep}$} & \colhead{Warning} \\
\colhead{} & \colhead{} & \colhead{(km s$^{-1}$)} &
\colhead{(km s$^{-1}$)} & \colhead{} & \colhead{} &
\colhead{} & \colhead{}
}
\startdata
5975817629354504704 & G352.8 & -57.0 & 11.1 & Gaia & Bary & 14 & \nodata \\
5975818350909013248 & G352.8 & -1.6 & 26.1 & Gaia & Bary & 11 & G \\
5975818831945360640 & G352.8 & 143.0 & 11.5 & Gaia & Bary & 11 & \nodata \\
5976050382223359744 & P24 & 8.4 & 6.0 & Gaia & Bary & 10 & \nodata \\
5976051035058690816 & P24 & 56.6 & 5.8 & Gaia & Bary & 12 & \nodata \\
5976051065119174144 & P24 & 9.8 & 8.1 & KMOS & U & 1 & B \\
5976051103777876736 & P24 & 290.6 & 16.8 & Gaia & Bary & 13 & \nodata \\
5976051138137614336 & P24 & 10.7 & 8.3 & KMOS & U & 1 & B \\
5976051271277228160 & P24 & 4.0 & 11.0 & KMOS & U & 1 & B \\
5976051271277547136 & P24 & -73.6 & 12.1 & KMOS & U & 1 & B \\
5976052065850557952 & P24 & 8.6 & 21.4 & Gaia & Bary & 13 & \nodata \\
5976057009362077184 & P24 & 21.3 & 5.8 & Gaia & Bary & 14 & \nodata \\
5976057013652868864 & P24 & -7.1 & 13.2 & KMOS & U & 1 & B \\
5976057078073012224 & P24 & 35.6 & 10.1 & KMOS & U & 1 & B \\
5976057082372353536 & P24 & -140.0 & 13.7 & Gaia & Bary & 13 & \nodata \\
5976057151091833088 & P24 & 0.7 & 11.0 & Gaia & Bary & 10 & \nodata \\
5976057941365819520 & P24 & -7.0 & 4.0 & KMOS & U & 1 & B \\
5976058491121932416 & P24 & 208.1 & 9.0 & Gaia & Bary & 16 & \nodata \\
5975845662604539008 & G353.1 & 28.0 & 8.4 & Gaia & Bary & 5 & \nodata \\
5975851228892454912 & G353.1 & -70.1 & 12.4 & KMOS & U & 1 & B \\
5975851542416457088 & G353.1 & 554.1 & 7.9 & Gaia & Bary & 9 & \nodata \\
5975851542416458752 & G353.1 & -508.0 & 5.6 & Gaia & Bary & 9 & \nodata \\
5976033339793075200 & G353.1 & 4.3 & 4.7 & Gaia & Bary & 8 & \nodata \\
5976039245374950400 & G353.1 & -34.8 & 11.5 & Gaia & Bary & 6 & \nodata \\
5975851851654317184 & G353.2 & -24.8 & 8.6 & Gaia & Bary & 8 & \nodata \\
5975851989093271936 & G353.2 & 109.9 & 9.3 & Gaia & Bary & 13 & \nodata \\
5975851989093271936 & G353.2 & 53.9 & 7.1 & KMOS & U & 1 & B \\
5975852092172488704 & G353.2 & 12.7 & 4.1 & KMOS & U & 1 & B \\
5975853466562022912 & G353.2 & 14.6 & 6.4 & Gaia & Bary & 14 & \nodata \\
5975853672720455936 & G353.2 & 74.8 & 9.0 & KMOS & U & 1 & B \\
5975853741439934336 & G353.2 & 0.5 & 2.4 & KMOS & U & 1 & B \\
5975853741439934976 & G353.2 & 206.7 & 12.4 & Gaia & Bary & 8 & \nodata \\
5976041551770860928 & G353.2 & -69.1 & 12.1 & Gaia & Bary & 6 & \nodata \\
5976041586130598656 & G353.2 & 16.5 & 13.1 & KMOS & U & 1 & B \\
5976041586130599552 & G353.2 & 67.9 & 7.4 & Gaia & Bary & 11 & \nodata \\
5976041826648771968 & G353.2 & 9.3 & 6.5 & Gaia & Bary & 7 & \nodata \\
5976041959788442112 & G353.2 & 38.1 & 4.4 & KMOS & U & 1 & B \\
5976041964087723904 & G353.2 & 76.1 & 14.4 & KMOS & U & 1 & B
\enddata
\tablecomments{
The table contains 38 records for 37 unique stars: 23 Gaia DR3
catalog aggregates and 15 KMOS measurements, with one star present
in both sources. P24 denotes Pismis~24; G352.8, G353.1, and G353.2
denote G352.8+0.6, G353.1+0.6, and G353.2+0.6, respectively.
$\sigma_{\rm RV}$ is the reported formal RV uncertainty.
Gaia values are barycentric catalog aggregates \citep{Katz2023},
denoted Bary in the Frame column. KMOS values are from Table B.2 of
\citet{RamirezTannus2021}; U indicates that the velocity frame is not
explicitly stated there. $n_{\rm rep}$ records catalog-reported transits
or visits, rather than independently verified dated RV measurements.
B denotes the literature binary/RV caution associated with the KMOS
photospheric measurements; it is not a confirmed binary classification
for every listed star. G denotes a Gaia RV variability indicator.
An empty warning entry provides no established diagnosis of RV
stability. Source 5975851989093271936 has separate Gaia and KMOS
entries. The individual records are retained for the reliability
assessment in this appendix; none is adopted as a validated tracer
of a cluster systemic velocity.
}
\end{deluxetable*}

The reported velocities show large star-to-star scatter within each cluster. For G352.8+0.6, the three Gaia DR3 measurements are $-57.0$, $-1.6$, and $+143.0$ km s$^{-1}$. Similarly large spreads occur in the other clusters: the Gaia DR3 measurements alone span approximately $-140$ to $+291$ km s$^{-1}$ in Pismis~24, $-508$ to $+554$ km s$^{-1}$ in G353.1+0.6, and $-69$ to $+207$ km s$^{-1}$ in G353.2+0.6. These discrepancies are present within a single catalog and reference frame and therefore cannot be attributed solely to differences between the Gaia and KMOS velocity conventions. In several cases, the differences between stars are also much larger than their reported formal RV uncertainties.

The available warning information does not provide a secure basis for selecting a subset of systemic-velocity tracers. The compilation identifies 8, 5, 7, and 3 stars with reported multiple visits or transits in Pismis~24, G353.1+0.6, G353.2+0.6, and G352.8+0.6, respectively. Binary/RV cautions are recorded for 7, 1, 7, and 0 stars, respectively, and one G352.8+0.6 member has a Gaia variability indicator. These warning counts are not counts of confirmed binaries or of stars known to have stable center-of-mass velocities. Conversely, the absence of a warning does not demonstrate RV stability. Orbital motion in unresolved binaries and other forms of stellar RV variability could contribute to the observed discrepancies, but the available temporal information is insufficient to determine the origin of each discrepant measurement. We therefore cannot define a defensible subset of stars from which to estimate a reliable systemic velocity for any of the four clusters.

Measurements from different surveys may additionally differ in velocity reference frame and instrumental or analysis zero point. Reference-frame differences can be converted when their definitions are known, whereas instrumental offsets require a separate calibration or assessment. In particular, the Gaia stellar velocities are reported in the barycentric frame, whereas the molecular-gas velocities used in this work are expressed as $V_{\rm LSR}$, and the reference frame of the tabulated KMOS measurements is not explicitly stated. We therefore do not compare the absolute stellar RVs directly with the molecular-gas $V_{\rm LSR}$ values. The very large star-to-star scatter among the Gaia DR3 measurements, however, is independent of this stellar--gas reference-frame issue and is itself sufficient to show that the available individual catalog RVs cannot be interpreted straightforwardly as tracers of cluster bulk motion.

Because the available stellar RV measurements do not permit reliable systemic velocities to be determined for the four clusters, we do not incorporate stellar radial velocities into the traceback analysis. The kinematic analysis is therefore restricted to the plane of the sky, as described in Sections~\ref{subsec:traceback} and \ref{subsec:results-kinematics}. Reliable cluster systemic velocities will require multi-epoch spectroscopy sufficient to identify RV-variable systems and constrain stellar center-of-mass velocities.

\section{Extended Molecular Gas Distribution toward NGC~6334 and NGC~6357}
\label{app:extended-co}

To place the molecular-gas structures identified toward NGC~6357 in a
broader spatial context, we examined the $^{12}$CO($J=1$--0) emission
over a field including NGC~6334, NGC~6357, and the region between the
two massive star-forming complexes.

The wide-field maps use the same local parent cube,
\texttt{dr6.s348.12co.fits}. The requested extraction covers
$351^\circ\leq l\leq354^\circ$ and $0^\circ\leq b\leq1.2^\circ$;
the actual retained pixel centers span
$351.004^\circ\leq l\leq353.999^\circ$ and
$0.002^\circ\leq b\leq1.029^\circ$. The data do not cover the full
requested upper-latitude limit. The beam is approximately $72\arcsec$,
the pixels are $24\arcsec$, and the channel spacing is $0.35048~\kms$.
The cube inherits four-channel spectral binning and intensity division
by 0.55. All 22 and 13
channels selected by the two integration intervals are available,
with channel centers from $-7.67587$ to $-0.31573~\kms$ and from
$+0.03475$ to $+4.24054~\kms$, respectively. We extracted native
pixels without additional mosaicking, reprojection, spatial smoothing,
spectral smoothing, or efficiency correction. Each map retains only
pixels finite in all channels of its integration interval.

We constructed the wide-field integrated-intensity maps using the same
velocity intervals adopted for the NGC~6357 analysis in
Section~\ref{subsec:gas-analysis}:
$-8.00\leq V_{\rm LSR}<-0.31~\kms$ and
$-0.31\leq V_{\rm LSR}<+4.25~\kms$.
These intervals correspond to the velocity ranges assigned to the
$-4$ and $+2~\kms$ components, respectively, in the principal
NGC~6357 analysis.

\begin{figure*}[!tp]
\centering
\includegraphics[width=\textwidth]{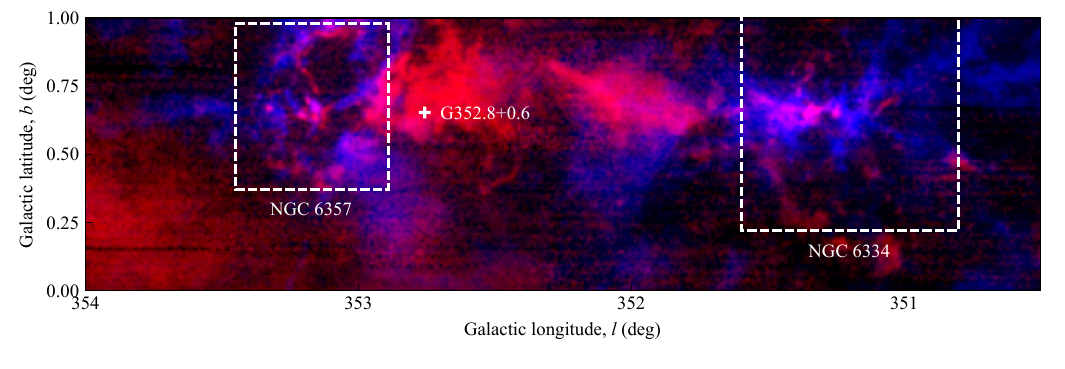}
\caption{
Two-color composite of the wide-field ThrUMMS
$^{12}$CO($J=1$--0) integrated-intensity distributions toward
NGC~6334 and NGC~6357. Blue shows emission integrated over
$-8.00\leq V_{\rm LSR}<-0.31~\kms$, and red shows emission integrated
over $-0.31\leq V_{\rm LSR}<+4.25~\kms$; these are the intervals
assigned to the $-4$ and $+2~\kms$ components toward NGC~6357.
Magenta regions show their projected overlap. The two color channels
are independently normalized and stretched to display the gas
morphology. White crosses and leader lines identify the reference
positions of NGC~6357 and NGC~6334 and the center of G352.8+0.6.
The image covers approximately $351^\circ\leq l\leq354^\circ$ and
$0^\circ\leq b\leq1^\circ$, with a $72\arcsec$ beam. The maps use
native pixels without additional spatial smoothing or reprojection;
pixels missing any required integration channel are masked.
}
\label{fig:app-extended-co}
\end{figure*}

Figure~\ref{fig:app-extended-co} combines the resulting wide-field
distributions in a red--blue image. Each color channel is normalized
independently to its positive-intensity 99.5th percentile and displayed
with an asinh stretch for the morphological comparison.

Emission within the velocity interval assigned to the $+2~\kms$
component toward NGC~6357 is also prominent toward the central region
of the neighboring massive star-forming complex NGC~6334. Emission in
both the $-4$ and $+2~\kms$ velocity intervals is also distributed over
the region between NGC~6334 and NGC~6357. On these larger spatial
scales, the two velocity-selected distributions show complementary
morphology in several regions rather than being confined to the immediate
vicinity of NGC~6357.

The emission toward NGC~6334 and across the intervening region suggests
that the $+2~\kms$ gas toward G352.8+0.6 may form part of a larger-scale
molecular environment associated with the NGC~6334--NGC~6357 complex.
This provides a broader spatial context for the molecular material
discussed in Section~\ref{subsubsec:plus2-association}.

\section{Analysis Settings and Numerical Implementation}
\label{app:analysis-details}

This appendix records the detailed settings for the methods described
in Section~\ref{sec:analysis}.

\subsection{Gaia HDBSCAN Settings}
\label{app:analysis-hdbscan}

We applied the Python \texttt{hdbscan} implementation
\citep{Campello2015,McInnes2017} to the five-dimensional Gaia astrometric
sample. Each coordinate was standardized over the full input sample as
\begin{equation}
z=\frac{x-\bar{x}}{s},
\label{eq:standardization}
\end{equation}
where $\bar{x}$ and $s$ are its mean and standard deviation. We used
Euclidean distance, \texttt{min\_cluster\_size}=10,
\texttt{min\_samples}=10, and the \texttt{leaf} selection method. The
minimum cluster size was an empirical choice of twice the number of
dimensions.

The saved member counts agree with the corresponding full-field
HDBSCAN labels.

\subsection{Distance, Isochrone, and Extinction Settings}
\label{app:analysis-cmd}

The statistical intervals were obtained from 20,000 bootstrap resamplings
of each fixed member list, preserving its size and recomputing the median
corrected parallax and its reciprocal in each realization (random seed
1864). No RUWE cut or additional Gaussian perturbation of the parallaxes
was applied. We report the 16th--84th percentiles. These intervals
characterize member-resampling uncertainty and do not include a separate
propagation of individual parallax measurement errors or residual Gaia
systematics. Residual systematics are considered in
Section~\ref{subsubsec:distance} and Appendix~\ref{app:parallax}.

The isochrone file was generated with CMD~3.9 and uses PARSEC v1.2S
with COLIBRI S\_37 + S\_35 + PR16, $Z_{\rm ini}=0.0152$ and
$[\mathrm{M/H}]=+0.01508$, the Gaia EDR3 Vegamag photometric system,
and YBC bolometric corrections \citep{Chen2019}. Extinction in the
three Gaia bands was evaluated with \texttt{dustapprox} version 0.1
\citep{Fouesneau2022}, using its precomputed polynomials based on the
Kurucz ODFNEW/NOVER atmosphere grid \citep{CastelliKurucz2003} and
the extinction law of \citet{Fitzpatrick1999} with $R_0=3.1$.
The degree-three model predicts $A_X/A_0$ as a function of $A_0$
and $T_{\rm eff}/5040$ for the \texttt{GAIA\_GAIA3.G},
\texttt{GAIA\_GAIA3.Gbp}, and \texttt{GAIA\_GAIA3.Grp} passbands;
model temperatures were clipped to 3500--50,000~K.

The grid covered
$6.0\leq\log_{10}(\mathrm{Age/yr})\leq8.0$ in 0.1~dex steps and
$0.50\leq A_0\leq12.0$~mag at 200 equally spaced values, with spacing
$11.5/199\simeq0.05779$~mag.

Because $G$ and $G_{\rm BP}-G_{\rm RP}$ span different numerical ranges,
we standardized each cluster's observed CMD using the member means and
population standard deviations (\texttt{ddof}=0). The same fixed
transformation was applied to every trial isochrone for that cluster.
All 257 selected members have finite photometry in the three bands and
were retained. The isochrones were interpolated along arc length in the
three-band magnitude space to at least 500 points or five times the
original point count, whichever is larger.

Doubling the curve sampling in
the full G353.1+0.6 grid leaves its best grid age unchanged and changes
its marginalized median log age and $A_0$ by only about
$2\times10^{-5}$~dex and $2\times10^{-4}$~mag, respectively.

The observed CMD scatter exceeds the median color uncertainty of
approximately 0.01~mag. We used the observed residual dispersion as an
empirical measure of the scatter around each trial isochrone. It may
include differential extinction and other star-to-star variations not
modeled separately. For each trial curve, $\sigma_r$ is the sample
standard deviation of the member residuals (\texttt{ddof}=1), and
\begin{equation}
P_i=\frac{1}{\sqrt{2\pi\sigma_r^2}}
\exp\left(-\frac{r_i^2}{2\sigma_r^2}\right),
\label{eq:isochrone-probability}
\end{equation}
\begin{equation}
\mathcal{L}=\prod_i P_i.
\label{eq:isochrone-likelihood}
\end{equation}
This likelihood provides an empirical measure of relative fit quality.

Using uniform grid weighting in log age and $A_0$, we marginalized
$\mathcal{L}$ over the other parameter and normalized each distribution.
Percentiles were computed from trapezoidally integrated cumulative
distributions; log-age percentiles were then transformed to ages. We
report the 50th percentiles and 16th--84th percentile intervals. These
intervals characterize the adopted empirical fitting distribution and
exclude systematic uncertainties in the CMD distance, stellar models,
extinction prescription, and other unmodeled scatter.

\subsection{Completeness Thresholds and IMF Normalization}
\label{app:analysis-mass}

Mass assignment used the marginalized median log age rounded to the
nearest available isochrone age, median $A_0$, and fixed CMD distance.
Initial masses were taken from the PARSEC \texttt{Mini} column rather
than its present-day \texttt{Mass} column.

The adopted $G_{\rm comp}$ values were selected by inspecting the
member magnitude histograms. They are 18.5, 18.0, 17.5, and 17.5~mag
for Pismis~24, G353.1+0.6, G353.2+0.6, and G352.8+0.6,
respectively. $M_{\rm comp}$ is the minimum initial mass among sampled
isochrone points brighter than or equal to $G_{\rm comp}$; the resulting
values are 1.85, 2.16, 2.36, and $1.64~M_\odot$, respectively.

For dimensionless initial mass $m=M/M_\odot$, we adopted the Kroupa IMF
\citep{Kroupa2001} with the unnormalized form
\begin{equation}
\phi(m)=
\begin{cases}
m^{-1.3}, & 0.08\leq m<0.5,\\
0.5\,m^{-2.3}, & 0.5\leq m\leq100.
\end{cases}
\label{eq:kroupa-imf}
\end{equation}
We write the corresponding mass distribution as
$dN/dm=C\phi(m)$ and determine $C$ from the observed members above
$m_{\rm comp}=M_{\rm comp}/M_\odot$.

The normalization upper limit was
$m_u=\min[100,m_{\rm surv,max}(t)]$, where $m_{\rm surv,max}(t)$ is the
maximum initial mass still represented by the adopted PARSEC isochrone at
the cluster age, including post-main-sequence phases. With $N_{\rm norm}$
denoting the selected members satisfying
$m_{\rm comp}\leq m\leq m_u$, we calculated
\begin{equation}
C=
\frac{N_{\rm norm}}
{\displaystyle\int_{m_{\rm comp}}^{m_u}\phi(m)\,dm},
\label{eq:imf-normalization}
\end{equation}
and
\begin{equation}
M_{\rm ini}
=
C\,M_\odot
\int_{0.08}^{100}m\,\phi(m)\,dm.
\label{eq:initial-cluster-mass}
\end{equation}

For G352.8+0.6, the maximum represented initial mass at the adopted
isochrone age is $23.86~M_\odot$, below the $300~M_\odot$ endpoint of
the youngest grid. The other three isochrones extend above
$100~M_\odot$, so their normalization is capped at $m_u=100$.
The normalization counts are 77, 24, 38, and 62 for Pismis~24,
G353.1+0.6, G353.2+0.6, and G352.8+0.6, respectively.

\subsection{Projection and Traceback Bootstrap}
\label{app:analysis-traceback}

Cluster centers were the median member right ascension and declination.
We used the equal-weight spherical mean of the three young-cluster
centers as the origin of a common gnomonic (TAN) projection, with axes
increasing east and north. Positions and proper-motion vectors were
transformed consistently onto this tangent plane.

Statistical uncertainties conditional on the adopted member lists were
estimated from 5000 bootstrap resamplings of the four member lists,
preserving their sizes and recalculating cluster centers and motions
(random seed 6357). No additional Gaussian astrometric perturbation was
applied. We evaluated the 50th and 16th--84th percentiles of the resulting
distributions.

\subsection{CO Spectrum and Position--Velocity Extraction}
\label{app:analysis-gas}

We constructed a mean CO spectrum over
$352.4^\circ\leq l\leq353.6^\circ$ and
$0.2^\circ\leq b\leq1.0^\circ$ and fitted three Gaussian components
with a linear baseline to characterize the principal spectral peaks.
The linear baseline was used only in fitting the mean spectrum;
the integrated-intensity maps were calculated directly from the CO cube.

The two integration intervals contain 22 and 13 channels, giving
$\sigma_W\simeq2.32$ and $1.79~\mathrm{K\,km\,s^{-1}}$ for the
$-4$ and $+2~\kms$ components, respectively. Contour significance was
therefore evaluated separately for the two velocity intervals.

The slice origin is $(l,b)=(352.75251^\circ,0.62753^\circ)$.
Within the longitude clipping region, its reference line runs from
$(353.667^\circ,0.83861^\circ)$ to
$(352.400^\circ,0.56214^\circ)$. In the local
$(\Delta l\cos b_0,\Delta b)$ plane, the transverse offsets extend
from $-0.15890^\circ$ to $+0.13637^\circ$, giving a full width of
$17.72\arcmin$. Position is the projection onto the relative-motion
direction, with positive offsets toward decreasing $l$ and $b$. Each
$24\arcsec$ position bin contains the equal-weight arithmetic mean of
finite native pixels within that bin and transverse range; no
interpolation or extra smoothing is applied.

\bibliographystyle{aasjournalv7.1}
\bibliography{references}

@ARTICLE{Barnes2015,
       author = {{Barnes}, Peter J. and {Muller}, Erik and {Indermuehle}, Balthasar and {O'Dougherty}, Stefan N. and {Lowe}, Vicki and {Cunningham}, Maria and {Hernandez}, Audra K. and {Fuller}, Gary A.},
        title = "{The Three-mm Ultimate Mopra Milky Way Survey. I. Survey Overview, Initial Data Releases, and First Results}",
      journal = {\apj},
         year = 2015,
        month = oct,
       volume = {812},
       number = {1},
          eid = {6},
        pages = {6},
          doi = {10.1088/0004-637X/812/1/6},
archivePrefix = {arXiv},
       eprint = {1507.05095},
 primaryClass = {astro-ph.GA},
       adsurl = {https://ui.adsabs.harvard.edu/abs/2015ApJ...812....6B}
}

@ARTICLE{Benjamin2003,
       author = {{Benjamin}, Robert A. and {Churchwell}, E. and {Babler}, Brian L. and {Bania}, T.~M. and {Clemens}, Dan P. and {Cohen}, Martin and {Dickey}, John M. and {Indebetouw}, R{\'e}my and {Jackson}, James M. and {Kobulnicky}, Henry A. and {Lazarian}, Alex and {Marston}, A.~P. and {Mathis}, John S. and {Meade}, Marilyn R. and {Seager}, Sara and {Stolovy}, S.~R. and {Watson}, C. and {Whitney}, Barbara A. and {Wolff}, Michael J. and {Wolfire}, Mark G.},
        title = "{GLIMPSE. I. An SIRTF Legacy Project to Map the Inner Galaxy}",
      journal = {\pasp},
         year = 2003,
        month = aug,
       volume = {115},
       number = {810},
        pages = {953-964},
          doi = {10.1086/376696},
archivePrefix = {arXiv},
       eprint = {astro-ph/0306274},
 primaryClass = {astro-ph},
       adsurl = {https://ui.adsabs.harvard.edu/abs/2003PASP..115..953B}
}

@ARTICLE{Bressan2012,
       author = {{Bressan}, Alessandro and {Marigo}, Paola and {Girardi}, L{\'e}o. and {Salasnich}, Bernardo and {Dal Cero}, Claudia and {Rubele}, Stefano and {Nanni}, Ambra},
        title = "{PARSEC: stellar tracks and isochrones with the PAdova and TRieste Stellar Evolution Code}",
      journal = {\mnras},
         year = 2012,
        month = nov,
       volume = {427},
       number = {1},
        pages = {127-145},
          doi = {10.1111/j.1365-2966.2012.21948.x},
archivePrefix = {arXiv},
       eprint = {1208.4498},
 primaryClass = {astro-ph.SR},
       adsurl = {https://ui.adsabs.harvard.edu/abs/2012MNRAS.427..127B}
}

@article{Campello2015,
author = {Campello, Ricardo J. G. B. and Moulavi, Davoud and Zimek, Arthur and Sander, J{\"o}rg},
title = {Hierarchical Density Estimates for Data Clustering, Visualization, and Outlier Detection},
year = {2015},
publisher = {Association for Computing Machinery},
address = {New York, NY, USA},
volume = {10},
number = {1},
issn = {1556-4681},
url = {https://doi.org/10.1145/2733381},
doi = {10.1145/2733381},
journal = {ACM Trans. Knowl. Discov. Data},
month = jul,
articleno = {5},
pages = {5},
numpages = {51},
}

@ARTICLE{Cappa2011,
       author = {{Cappa}, C.~E. and {Barb{\'a}}, R. and {Duronea}, N.~U. and {Vasquez}, J. and {Arnal}, E.~M. and {Goss}, W.~M. and {Fern{\'a}ndez Laj{\'u}s}, E.},
        title = "{A multifrequency study of the active star-forming complex NGC 6357 - I. Interstellar structures linked to the open cluster Pis 24}",
      journal = {\mnras},
         year = 2011,
        month = aug,
       volume = {415},
       number = {3},
        pages = {2844-2858},
          doi = {10.1111/j.1365-2966.2011.18902.x},
archivePrefix = {arXiv},
       eprint = {1104.2493},
 primaryClass = {astro-ph.GA},
       adsurl = {https://ui.adsabs.harvard.edu/abs/2011MNRAS.415.2844C}
}

@ARTICLE{Carey2009,
       author = {{Carey}, S.~J. and {Noriega-Crespo}, A. and {Mizuno}, D.~R. and {Shenoy}, S. and {Paladini}, R. and {Kraemer}, K.~E. and {Price}, S.~D. and {Flagey}, N. and {Ryan}, E. and {Ingalls}, J.~G. and {Kuchar}, T.~A. and {Pinheiro Gon{\c{c}}alves}, Daniela and {Indebetouw}, R. and {Billot}, N. and {Marleau}, F.~R. and {Padgett}, D.~L. and {Rebull}, L.~M. and {Bressert}, E. and {Ali}, Babar and {Molinari}, S. and {Martin}, P.~G. and {Berriman}, G.~B. and {Boulanger}, F. and {Latter}, W.~B. and {Miville-Deschenes}, M.~A. and {Shipman}, R. and {Testi}, L.},
        title = "{MIPSGAL: A Survey of the Inner Galactic Plane at 24 and 70 {\ensuremath{\mu}}m}",
      journal = {\pasp},
         year = 2009,
        month = jan,
       volume = {121},
       number = {875},
        pages = {76},
          doi = {10.1086/596581},
       adsurl = {https://ui.adsabs.harvard.edu/abs/2009PASP..121...76C}
}

@ARTICLE{Ahumada2007,
       author = {{Ahumada}, A.~V. and {Clari{\'a}}, J.~J. and {Bica}, E.},
        title = "{Integrated spectral properties of 22 small angular diameter galactic open clusters}",
      journal = {\aap},
         year = 2007,
        month = oct,
       volume = {473},
       number = {2},
        pages = {437-444},
          doi = {10.1051/0004-6361:20077608},
       adsurl = {https://ui.adsabs.harvard.edu/abs/2007A&A...473..437A}
}

@ARTICLE{Lima2014,
       author = {{Lima}, E.~F. and {Bica}, E. and {Bonatto}, C. and {Saito}, R.~K.},
        title = "{Probing embedded star clusters in the HII complex NGC 6357 with VVV}",
      journal = {\aap},
         year = 2014,
        month = aug,
       volume = {568},
          eid = {A16},
        pages = {A16},
          doi = {10.1051/0004-6361/201323050},
archivePrefix = {arXiv},
       eprint = {1406.2413},
 primaryClass = {astro-ph.SR},
       adsurl = {https://ui.adsabs.harvard.edu/abs/2014A&A...568A..16L}
}

@ARTICLE{Chen2014,
       author = {{Chen}, Yang and {Girardi}, L{\'e}o and {Bressan}, Alessandro and {Marigo}, Paola and {Barbieri}, Mauro and {Kong}, Xu},
        title = "{Improving PARSEC models for very low mass stars}",
      journal = {\mnras},
         year = 2014,
        month = nov,
       volume = {444},
       number = {3},
        pages = {2525-2543},
          doi = {10.1093/mnras/stu1605},
archivePrefix = {arXiv},
       eprint = {1409.0322},
 primaryClass = {astro-ph.SR},
       adsurl = {https://ui.adsabs.harvard.edu/abs/2014MNRAS.444.2525C}
}

@ARTICLE{Fitzpatrick1999,
       author = {{Fitzpatrick}, Edward L.},
        title = "{Correcting for the Effects of Interstellar Extinction}",
      journal = {\pasp},
         year = 1999,
        month = jan,
       volume = {111},
       number = {755},
        pages = {63-75},
          doi = {10.1086/316293},
archivePrefix = {arXiv},
       eprint = {astro-ph/9809387},
 primaryClass = {astro-ph},
       adsurl = {https://ui.adsabs.harvard.edu/abs/1999PASP..111...63F}
}

@ARTICLE{Chen2019,
       author = {{Chen}, Yang and {Girardi}, L{\'e}o and {Fu}, Xiaoting and {Bressan}, Alessandro and {Aringer}, Bernhard and {Dal Tio}, Piero and {Pastorelli}, Giada and {Marigo}, Paola and {Costa}, Guglielmo and {Zhang}, Xing},
        title = "{YBC: a stellar bolometric corrections database with variable extinction coefficients. Application to PARSEC isochrones}",
      journal = {\aap},
         year = 2019,
        month = dec,
       volume = {632},
          eid = {A105},
        pages = {A105},
          doi = {10.1051/0004-6361/201936612},
archivePrefix = {arXiv},
       eprint = {1910.09037},
 primaryClass = {astro-ph.SR},
       adsurl = {https://ui.adsabs.harvard.edu/abs/2019A&A...632A.105C}
}

@misc{Fouesneau2022,
  author = {Fouesneau, Morgan and Andrae, Ren{\'{e}} and Sordo, Rosanna and Dharmawardena, Thavisha},
  title = {{dustapprox}},
  year = {2022},
  version = {0.1},
  howpublished = {Software},
  url = {https://github.com/mfouesneau/dustapprox},
}

@ARTICLE{Chen2015,
       author = {{Chen}, Yang and {Bressan}, Alessandro and {Girardi}, L{\'e}o and {Marigo}, Paola and {Kong}, Xu and {Lanza}, Antonio},
        title = "{PARSEC evolutionary tracks of massive stars up to 350 M$_{☉}$ at metallicities 0.0001 {\ensuremath{\leq}} Z {\ensuremath{\leq}} 0.04}",
      journal = {\mnras},
         year = 2015,
        month = sep,
       volume = {452},
       number = {1},
        pages = {1068-1080},
          doi = {10.1093/mnras/stv1281},
archivePrefix = {arXiv},
       eprint = {1506.01681},
 primaryClass = {astro-ph.SR},
       adsurl = {https://ui.adsabs.harvard.edu/abs/2015MNRAS.452.1068C}
}

@ARTICLE{Dale2015,
       author = {{Dale}, James E.},
        title = "{The modelling of feedback in star formation simulations}",
      journal = {\nar},
         year = 2015,
        month = oct,
       volume = {68},
        pages = {1-33},
          doi = {10.1016/j.newar.2015.06.001},
archivePrefix = {arXiv},
       eprint = {1508.06054},
 primaryClass = {astro-ph.GA},
       adsurl = {https://ui.adsabs.harvard.edu/abs/2015NewAR..68....1D}
}

@ARTICLE{Drew2014,
       author = {{Drew}, J.~E. and {Gonzalez-Solares}, E. and {Greimel}, R. and {Irwin}, M.~J. and {K{\"u}pc{\"u} Yoldas}, A. and {Lewis}, J. and {Barentsen}, G. and {Eisl{\"o}ffel}, J. and {Farnhill}, H.~J. and {Martin}, W.~E. and {Walsh}, J.~R. and {Walton}, N.~A. and {Mohr-Smith}, M. and {Raddi}, R. and {Sale}, S.~E. and {Wright}, N.~J. and {Groot}, P. and {Barlow}, M.~J. and {Corradi}, R.~L.~M. and {Drake}, J.~J. and {Fabregat}, J. and {Frew}, D.~J. and {G{\"a}nsicke}, B.~T. and {Knigge}, C. and {Mampaso}, A. and {Morris}, R.~A.~H. and {Naylor}, T. and {Parker}, Q.~A. and {Phillipps}, S. and {Ruhland}, C. and {Steeghs}, D. and {Unruh}, Y.~C. and {Vink}, J.~S. and {Wesson}, R. and {Zijlstra}, A.~A.},
        title = "{The VST Photometric H{\ensuremath{\alpha}} Survey of the Southern Galactic Plane and Bulge (VPHAS+)}",
      journal = {\mnras},
         year = 2014,
        month = may,
       volume = {440},
       number = {3},
        pages = {2036-2058},
          doi = {10.1093/mnras/stu394},
archivePrefix = {arXiv},
       eprint = {1402.7024},
 primaryClass = {astro-ph.GA},
       adsurl = {https://ui.adsabs.harvard.edu/abs/2014MNRAS.440.2036D}
}

@ARTICLE{Fang2012,
       author = {{Fang}, M. and {van Boekel}, R. and {King}, R.~R. and {Henning}, Th. and {Bouwman}, J. and {Doi}, Y. and {Okamoto}, Y.~K. and {Roccatagliata}, V. and {Sicilia-Aguilar}, A.},
        title = "{Star formation and disk properties in Pismis 24}",
      journal = {\aap},
         year = 2012,
        month = mar,
       volume = {539},
          eid = {A119},
        pages = {A119},
          doi = {10.1051/0004-6361/201015914},
archivePrefix = {arXiv},
       eprint = {1201.0833},
 primaryClass = {astro-ph.SR},
       adsurl = {https://ui.adsabs.harvard.edu/abs/2012A&A...539A.119F}
}

@ARTICLE{Ferreira2021,
       author = {{Ferreira}, F.~A. and {Corradi}, W.~J.~B. and {Maia}, F.~F.~S. and {Angelo}, M.~S. and {Santos}, Jr., J.~F.~C.},
        title = "{New star clusters discovered towards the Galactic bulge direction using Gaia DR2}",
      journal = {\mnras},
         year = 2021,
        month = mar,
       volume = {502},
       number = {1},
        pages = {L90-L94},
          doi = {10.1093/mnrasl/slab011},
archivePrefix = {arXiv},
       eprint = {2101.10982},
 primaryClass = {astro-ph.GA},
       adsurl = {https://ui.adsabs.harvard.edu/abs/2021MNRAS.502L..90F}
}

@ARTICLE{GaiaDR3,
       author = {{Gaia Collaboration} and {Vallenari}, A. and {Brown}, A.~G.~A. and {Prusti}, T. and {de Bruijne}, J.~H.~J. and {Arenou}, F. and {Babusiaux}, C. and {Biermann}, M. and {Creevey}, O.~L. and {Ducourant}, C. and {Evans}, D.~W. and {Eyer}, L. and {Guerra}, R. and {Hutton}, A. and {Jordi}, C. and {Klioner}, S.~A. and {Lammers}, U.~L. and {Lindegren}, L. and {Luri}, X. and {Mignard}, F. and {Panem}, C. and {Pourbaix}, D. and {Randich}, S. and {Sartoretti}, P. and {Soubiran}, C. and {Tanga}, P. and {Walton}, N.~A. and {Bailer-Jones}, C.~A.~L. and {Bastian}, U. and {Drimmel}, R. and {Jansen}, F. and {Katz}, D. and {Lattanzi}, M.~G. and {van Leeuwen}, F. and {Bakker}, J. and {Cacciari}, C. and {Casta{\~n}eda}, J. and {De Angeli}, F. and {Fabricius}, C. and {Fouesneau}, M. and {Fr{\'e}mat}, Y. and {Galluccio}, L. and {Guerrier}, A. and {Heiter}, U. and {Masana}, E. and {Messineo}, R. and {Mowlavi}, N. and {Nicolas}, C. and {Nienartowicz}, K. and {Pailler}, F. and {Panuzzo}, P. and {Riclet}, F. and {Roux}, W. and {Seabroke}, G.~M. and {Sordo}, R. and {Th{\'e}venin}, F. and {Gracia-Abril}, G. and {Portell}, J. and {Teyssier}, D. and {Altmann}, M. and {Andrae}, R. and {Audard}, M. and {Bellas-Velidis}, I. and {Benson}, K. and {Berthier}, J. and {Blomme}, R. and {Burgess}, P.~W. and {Busonero}, D. and {Busso}, G. and {C{\'a}novas}, H. and {Carry}, B. and {Cellino}, A. and {Cheek}, N. and {Clementini}, G. and {Damerdji}, Y. and {Davidson}, M. and {de Teodoro}, P. and {Nu{\~n}ez Campos}, M. and {Delchambre}, L. and {Dell'Oro}, A. and {Esquej}, P. and {Fern{\'a}ndez-Hern{\'a}ndez}, J. and {Fraile}, E. and {Garabato}, D. and {Garc{\'\i}a-Lario}, P. and {Gosset}, E. and {Haigron}, R. and {Halbwachs}, J.-L. and {Hambly}, N.~C. and {Harrison}, D.~L. and {Hern{\'a}ndez}, J. and {Hestroffer}, D. and {Hodgkin}, S.~T. and {Holl}, B. and {Jan{\ss}en}, K. and {Jevardat de Fombelle}, G. and {Jordan}, S. and {Krone-Martins}, A. and {Lanzafame}, A.~C. and {L{\"o}ffler}, W. and {Marchal}, O. and {Marrese}, P.~M. and {Moitinho}, A. and {Muinonen}, K. and {Osborne}, P. and {Pancino}, E. and {Pauwels}, T. and {Recio-Blanco}, A. and {Reyl{\'e}}, C. and {Riello}, M. and {Rimoldini}, L. and {Roegiers}, T. and {Rybizki}, J. and {Sarro}, L.~M. and {Siopis}, C. and {Smith}, M. and {Sozzetti}, A. and {Utrilla}, E. and {van Leeuwen}, M. and {Abbas}, U. and {{\'A}brah{\'a}m}, P. and {Abreu Aramburu}, A. and {Aerts}, C. and {Aguado}, J.~J. and {Ajaj}, M. and {Aldea-Montero}, F. and {Altavilla}, G. and {{\'A}lvarez}, M.~A. and {Alves}, J. and {Anders}, F. and {Anderson}, R.~I. and {Anglada Varela}, E. and {Antoja}, T. and {Baines}, D. and {Baker}, S.~G. and {Balaguer-N{\'u}{\~n}ez}, L. and {Balbinot}, E. and {Balog}, Z. and {Barache}, C. and {Barbato}, D. and {Barros}, M. and {Barstow}, M.~A. and {Bartolom{\'e}}, S. and {Bassilana}, J.-L. and {Bauchet}, N. and {Becciani}, U. and {Bellazzini}, M. and {Berihuete}, A. and {Bernet}, M. and {Bertone}, S. and {Bianchi}, L. and {Binnenfeld}, A. and {Blanco-Cuaresma}, S. and {Blazere}, A. and {Boch}, T. and {Bombrun}, A. and {Bossini}, D. and {Bouquillon}, S. and {Bragaglia}, A. and {Bramante}, L. and {Breedt}, E. and {Bressan}, A. and {Brouillet}, N. and {Brugaletta}, E. and {Bucciarelli}, B. and {Burlacu}, A. and {Butkevich}, A.~G. and {Buzzi}, R. and {Caffau}, E. and {Cancelliere}, R. and {Cantat-Gaudin}, T. and {Carballo}, R. and {Carlucci}, T. and {Carnerero}, M.~I. and {Carrasco}, J.~M. and {Casamiquela}, L. and {Castellani}, M. and {Castro-Ginard}, A. and {Chaoul}, L. and {Charlot}, P. and {Chemin}, L. and {Chiaramida}, V. and {Chiavassa}, A. and {Chornay}, N. and {Comoretto}, G. and {Contursi}, G. and {Cooper}, W.~J. and {Cornez}, T. and {Cowell}, S. and {Crifo}, F. and {Cropper}, M. and {Crosta}, M. and {Crowley}, C. and {Dafonte}, C. and {Dapergolas}, A. and {David}, M. and {David}, P. and {de Laverny}, P. and {De Luise}, F. and {De March}, R.},
        title = "{Gaia Data Release 3. Summary of the content and survey properties}",
      journal = {\aap},
         year = 2023,
        month = jun,
       volume = {674},
          eid = {A1},
        pages = {A1},
          doi = {10.1051/0004-6361/202243940},
archivePrefix = {arXiv},
       eprint = {2208.00211},
 primaryClass = {astro-ph.GA},
       adsurl = {https://ui.adsabs.harvard.edu/abs/2023A&A...674A...1G}
}

@ARTICLE{Haisch2001,
       author = {{Haisch}, Jr., Karl E. and {Lada}, Elizabeth A. and {Lada}, Charles J.},
        title = "{Disk Frequencies and Lifetimes in Young Clusters}",
      journal = {\apjl},
         year = 2001,
        month = jun,
       volume = {553},
       number = {2},
        pages = {L153-L156},
          doi = {10.1086/320685},
archivePrefix = {arXiv},
       eprint = {astro-ph/0104347},
 primaryClass = {astro-ph},
       adsurl = {https://ui.adsabs.harvard.edu/abs/2001ApJ...553L.153H}
}

@ARTICLE{HuntReffert2021,
       author = {{Hunt}, Emily L. and {Reffert}, Sabine},
        title = "{Improving the open cluster census. I. Comparison of clustering algorithms applied to Gaia DR2 data}",
      journal = {\aap},
         year = 2021,
        month = feb,
       volume = {646},
          eid = {A104},
        pages = {A104},
          doi = {10.1051/0004-6361/202039341},
archivePrefix = {arXiv},
       eprint = {2012.04267},
 primaryClass = {astro-ph.GA},
       adsurl = {https://ui.adsabs.harvard.edu/abs/2021A&A...646A.104H}
}

@ARTICLE{HuntReffert2023,
       author = {{Hunt}, Emily L. and {Reffert}, Sabine},
        title = "{Improving the open cluster census. II. An all-sky cluster catalogue with Gaia DR3}",
      journal = {\aap},
         year = 2023,
        month = may,
       volume = {673},
          eid = {A114},
        pages = {A114},
          doi = {10.1051/0004-6361/202346285},
archivePrefix = {arXiv},
       eprint = {2303.13424},
 primaryClass = {astro-ph.GA},
       adsurl = {https://ui.adsabs.harvard.edu/abs/2023A&A...673A.114H}
}

@ARTICLE{Katz2023,
       author = {{Katz}, D. and {Sartoretti}, P. and {Guerrier}, A. and {Panuzzo}, P. and {Seabroke}, G.~M. and {Th{\'e}venin}, F. and {Cropper}, M. and {Benson}, K. and {Blomme}, R. and {Haigron}, R. and {Marchal}, O. and {Smith}, M. and {Baker}, S. and {Chemin}, L. and {Damerdji}, Y. and {David}, M. and {Dolding}, C. and {Fr{\'e}mat}, Y. and {Gosset}, E. and {Jan{\ss}en}, K. and {Jasniewicz}, G. and {Lobel}, A. and {Plum}, G. and {Samaras}, N. and {Snaith}, O. and {Soubiran}, C. and {Vanel}, O. and {Zwitter}, T. and {Antoja}, T. and {Arenou}, F. and {Babusiaux}, C. and {Brouillet}, N. and {Caffau}, E. and {Di Matteo}, P. and {Fabre}, C. and {Fabricius}, C. and {Fragkoudi}, F. and {Haywood}, M. and {Huckle}, H.~E. and {Hottier}, C. and {Lasne}, Y. and {Leclerc}, N. and {Mastrobuono-Battisti}, A. and {Royer}, F. and {Teyssier}, D. and {Zorec}, J. and {Crifo}, F. and {Jean-Antoine Piccolo}, A. and {Turon}, C. and {Viala}, Y.},
        title = "{Gaia Data Release 3. Properties and validation of the radial velocities}",
      journal = {\aap},
         year = 2023,
        month = jun,
       volume = {674},
          eid = {A5},
        pages = {A5},
          doi = {10.1051/0004-6361/202244220},
archivePrefix = {arXiv},
       eprint = {2206.05902},
 primaryClass = {astro-ph.GA},
       adsurl = {https://ui.adsabs.harvard.edu/abs/2023A&A...674A...5K}
}

@ARTICLE{Kroupa2001,
       author = {{Kroupa}, Pavel},
        title = "{On the variation of the initial mass function}",
      journal = {\mnras},
         year = 2001,
        month = apr,
       volume = {322},
       number = {2},
        pages = {231-246},
          doi = {10.1046/j.1365-8711.2001.04022.x},
archivePrefix = {arXiv},
       eprint = {astro-ph/0009005},
 primaryClass = {astro-ph},
       adsurl = {https://ui.adsabs.harvard.edu/abs/2001MNRAS.322..231K}
}

@INPROCEEDINGS{Krumholz2014,
       author = {{Krumholz}, M.~R. and {Bate}, M.~R. and {Arce}, H.~G. and {Dale}, J.~E. and {Gutermuth}, R. and {Klein}, R.~I. and {Li}, Z.-Y. and {Nakamura}, F. and {Zhang}, Q.},
        title = "{Star Cluster Formation and Feedback}",
    booktitle = {Protostars and Planets VI},
         year = 2014,
       editor = {{Beuther}, Henrik and {Klessen}, Ralf S. and {Dullemond}, Cornelis P. and {Henning}, Thomas},
        month = jan,
        pages = {243-266},
          doi = {10.2458/azu_uapress_9780816531240-ch011},
archivePrefix = {arXiv},
       eprint = {1401.2473},
 primaryClass = {astro-ph.GA},
       adsurl = {https://ui.adsabs.harvard.edu/abs/2014prpl.conf..243K}
}

@ARTICLE{Lindegren2021,
       author = {{Lindegren}, L. and {Bastian}, U. and {Biermann}, M. and {Bombrun}, A. and {de Torres}, A. and {Gerlach}, E. and {Geyer}, R. and {Hern{\'a}ndez}, J. and {Hilger}, T. and {Hobbs}, D. and {Klioner}, S.~A. and {Lammers}, U. and {McMillan}, P.~J. and {Ramos-Lerate}, M. and {Steidelm{\"u}ller}, H. and {Stephenson}, C.~A. and {van Leeuwen}, F.},
        title = "{Gaia Early Data Release 3. Parallax bias versus magnitude, colour, and position}",
      journal = {\aap},
         year = 2021,
        month = may,
       volume = {649},
          eid = {A4},
        pages = {A4},
          doi = {10.1051/0004-6361/202039653},
archivePrefix = {arXiv},
       eprint = {2012.01742},
 primaryClass = {astro-ph.IM},
       adsurl = {https://ui.adsabs.harvard.edu/abs/2021A&A...649A...4L}
}

@ARTICLE{Marigo2017,
       author = {{Marigo}, Paola and {Girardi}, L{\'e}o and {Bressan}, Alessandro and {Rosenfield}, Philip and {Aringer}, Bernhard and {Chen}, Yang and {Dussin}, Marco and {Nanni}, Ambra and {Pastorelli}, Giada and {Rodrigues}, Tha{\'\i}se S. and {Trabucchi}, Michele and {Bladh}, Sara and {Dalcanton}, Julianne and {Groenewegen}, Martin A.~T. and {Montalb{\'a}n}, Josefina and {Wood}, Peter R.},
        title = "{A New Generation of PARSEC-COLIBRI Stellar Isochrones Including the TP-AGB Phase}",
      journal = {\apj},
         year = 2017,
        month = jan,
       volume = {835},
       number = {1},
          eid = {77},
        pages = {77},
          doi = {10.3847/1538-4357/835/1/77},
archivePrefix = {arXiv},
       eprint = {1701.08510},
 primaryClass = {astro-ph.SR},
       adsurl = {https://ui.adsabs.harvard.edu/abs/2017ApJ...835...77M}
}

@ARTICLE{Massi2015,
       author = {{Massi}, F. and {Giannetti}, A. and {Di Carlo}, E. and {Brand}, J. and {Beltr{\'a}n}, M.~T. and {Marconi}, G.},
        title = "{Young open clusters in the Galactic star forming region NGC 6357}",
      journal = {\aap},
         year = 2015,
        month = jan,
       volume = {573},
          eid = {A95},
        pages = {A95},
          doi = {10.1051/0004-6361/201424388},
archivePrefix = {arXiv},
       eprint = {1410.4340},
 primaryClass = {astro-ph.SR},
       adsurl = {https://ui.adsabs.harvard.edu/abs/2015A&A...573A..95M}
}

@ARTICLE{McInnes2017,
       author = {{McInnes}, Leland and {Healy}, John and {Astels}, Steve},
        title = "{hdbscan: Hierarchical density based clustering}",
      journal = {The Journal of Open Source Software},
         year = 2017,
        month = mar,
       volume = {2},
       number = {11},
          eid = {205},
        pages = {205},
          doi = {10.21105/joss.00205},
       adsurl = {https://ui.adsabs.harvard.edu/abs/2017JOSS....2..205M}
}

@ARTICLE{Pastorelli2019,
       author = {{Pastorelli}, Giada and {Marigo}, Paola and {Girardi}, L{\'e}o and {Chen}, Yang and {Rubele}, Stefano and {Trabucchi}, Michele and {Aringer}, Bernhard and {Bladh}, Sara and {Bressan}, Alessandro and {Montalb{\'a}n}, Josefina and {Boyer}, Martha L. and {Dalcanton}, Julianne J. and {Eriksson}, Kjell and {Groenewegen}, Martin A.~T. and {H{\"o}fner}, Susanne and {Lebzelter}, Thomas and {Nanni}, Ambra and {Rosenfield}, Philip and {Wood}, Peter R. and {Cioni}, Maria-Rosa L.},
        title = "{Constraining the thermally pulsing asymptotic giant branch phase with resolved stellar populations in the Small Magellanic Cloud}",
      journal = {\mnras},
         year = 2019,
        month = jun,
       volume = {485},
       number = {4},
        pages = {5666-5692},
          doi = {10.1093/mnras/stz725},
archivePrefix = {arXiv},
       eprint = {1903.04499},
 primaryClass = {astro-ph.SR},
       adsurl = {https://ui.adsabs.harvard.edu/abs/2019MNRAS.485.5666P}
}

@ARTICLE{Pastorelli2020,
       author = {{Pastorelli}, Giada and {Marigo}, Paola and {Girardi}, L{\'e}o and {Aringer}, Bernhard and {Chen}, Yang and {Rubele}, Stefano and {Trabucchi}, Michele and {Bladh}, Sara and {Boyer}, Martha L. and {Bressan}, Alessandro and {Dalcanton}, Julianne J. and {Groenewegen}, Martin A.~T. and {Lebzelter}, Thomas and {Mowlavi}, Nami and {Chubb}, Katy L. and {Cioni}, Maria-Rosa L. and {de Grijs}, Richard and {Ivanov}, Valentin D. and {Nanni}, Ambra and {van Loon}, Jacco Th and {Zaggia}, Simone},
        title = "{Constraining the thermally pulsing asymptotic giant branch phase with resolved stellar populations in the Large Magellanic Cloud}",
      journal = {\mnras},
         year = 2020,
        month = nov,
       volume = {498},
       number = {3},
        pages = {3283-3301},
          doi = {10.1093/mnras/staa2565},
archivePrefix = {arXiv},
       eprint = {2008.08595},
 primaryClass = {astro-ph.SR},
       adsurl = {https://ui.adsabs.harvard.edu/abs/2020MNRAS.498.3283P}
}

@ARTICLE{Povich2026,
       author = {{Povich}, Matthew S. and {Townsley}, Leisa K. and {Broos}, Patrick S. and {Bonilla}, Aldair E. and {Nguyen}, Giaky and {Soos}, Carly and {Kuhn}, Michael A. and {Singh}, Simran S. and {Garmire}, Gordon P.},
        title = "{Structure and Large-Scale Kinematics of Young Stellar Populations in the NGC 6357 and NGC 6334 Giant Molecular Cloud Complex}",
      journal = {arXiv e-prints},
         year = 2026,
        month = apr,
          eid = {arXiv:2604.15050},
        pages = {arXiv:2604.15050},
          doi = {10.48550/arXiv.2604.15050},
archivePrefix = {arXiv},
       eprint = {2604.15050},
 primaryClass = {astro-ph.GA},
       adsurl = {https://ui.adsabs.harvard.edu/abs/2026arXiv260415050P}
}

@ARTICLE{Preibisch2005,
       author = {{Preibisch}, Thomas and {Feigelson}, Eric D.},
        title = "{The Evolution of X-Ray Emission in Young Stars}",
      journal = {\apjs},
         year = 2005,
        month = oct,
       volume = {160},
       number = {2},
        pages = {390-400},
          doi = {10.1086/432094},
archivePrefix = {arXiv},
       eprint = {astro-ph/0506052},
 primaryClass = {astro-ph},
       adsurl = {https://ui.adsabs.harvard.edu/abs/2005ApJS..160..390P}
}

@ARTICLE{RamirezTannus2020,
       author = {{Ram{\'\i}rez-Tannus}, M.~C. and {Poorta}, J. and {Bik}, A. and {Kaper}, L. and {de Koter}, A. and {De Ridder}, J. and {Beuther}, H. and {Brandner}, W. and {Davies}, B. and {Gennaro}, M. and {Guo}, D. and {Henning}, T. and {Linz}, H. and {Naylor}, T. and {Pasquali}, A. and {Ram{\'\i}rez-Agudelo}, O.~H. and {Sana}, H.},
        title = "{The young stellar content of the giant H II regions M 8, G333.6-0.2, and NGC 6357 with VLT/KMOS}",
      journal = {\aap},
         year = 2020,
        month = jan,
       volume = {633},
          eid = {A155},
        pages = {A155},
          doi = {10.1051/0004-6361/201935941},
archivePrefix = {arXiv},
       eprint = {1911.09558},
 primaryClass = {astro-ph.SR},
       adsurl = {https://ui.adsabs.harvard.edu/abs/2020A&A...633A.155R}
}

@ARTICLE{RamirezTannus2021,
       author = {{Ram{\'\i}rez-Tannus}, M.~C. and {Backs}, F. and {de Koter}, A. and {Sana}, H. and {Beuther}, H. and {Bik}, A. and {Brandner}, W. and {Kaper}, L. and {Linz}, H. and {Henning}, Th. and {Poorta}, J.},
        title = "{A relation between the radial velocity dispersion of young clusters and their age. Evidence for hardening as the formation scenario of massive close binaries}",
      journal = {\aap},
         year = 2021,
        month = jan,
       volume = {645},
          eid = {L10},
        pages = {L10},
          doi = {10.1051/0004-6361/202039673},
archivePrefix = {arXiv},
       eprint = {2101.01604},
 primaryClass = {astro-ph.SR},
       adsurl = {https://ui.adsabs.harvard.edu/abs/2021A&A...645L..10R}
}

@ARTICLE{Russeil2017,
       author = {{Russeil}, D. and {Adami}, C. and {Bouret}, J.~C. and {Herv{\'e}}, A. and {Parker}, Q.~A. and {Zavagno}, A. and {Motte}, F.},
        title = "{NGC 6334 and NGC 6357. Insights from spectroscopy of their OB star populations}",
      journal = {\aap},
         year = 2017,
        month = nov,
       volume = {607},
          eid = {A86},
        pages = {A86},
          doi = {10.1051/0004-6361/201629870},
       adsurl = {https://ui.adsabs.harvard.edu/abs/2017A&A...607A..86R}
}

@ARTICLE{Russeil2019,
       author = {{Russeil}, D. and {Figueira}, M. and {Zavagno}, A. and {Motte}, F. and {Schneider}, N. and {Men'shchikov}, A. and {Bontemps}, S. and {Andr{\'e}}, P. and {Anderson}, L.~D. and {Benedettini}, M. and {Didelon}, P. and {Di Francesco}, J. and {Elia}, D. and {K{\"o}nyves}, V. and {Nguyen Luong}, Q. and {Nony}, T. and {Pezzuto}, S. and {Rygl}, K.~L.~J. and {Schisano}, E. and {Spinoglio}, L. and {Tig{\'e}}, J. and {White}, G.~J.},
        title = "{Herschel-HOBYS study of the earliest phases of high-mass star formation in NGC 6357}",
      journal = {\aap},
         year = 2019,
        month = may,
       volume = {625},
          eid = {A134},
        pages = {A134},
          doi = {10.1051/0004-6361/201833870},
       adsurl = {https://ui.adsabs.harvard.edu/abs/2019A&A...625A.134R}
}

@ARTICLE{Russeil2020,
       author = {{Russeil}, D. and {Zavagno}, A. and {Nguyen}, A. and {Figueira}, M. and {Adami}, C. and {Bouret}, J.~C.},
        title = "{OB stars and YSO populations in the region of NGC 6334-NGC 6357 as seen with Gaia DR2}",
      journal = {\aap},
         year = 2020,
        month = oct,
       volume = {642},
          eid = {A21},
        pages = {A21},
          doi = {10.1051/0004-6361/202037674},
       adsurl = {https://ui.adsabs.harvard.edu/abs/2020A&A...642A..21R}
}

@ARTICLE{Smith2025,
       author = {{Smith}, Leigh C. and {Lucas}, Philip W. and {Koposov}, Sergey E. and {Gonzalez-Fernandez}, Carlos and {Alonso-Garc{\'\i}a}, Javier and {Minniti}, Dante and {Sanders}, Jason L. and {Bedin}, Luigi R. and {Belokurov}, Vasily and {Evans}, N. Wyn and {Hempel}, Maren and {Ivanov}, Valentin D. and {Kurtev}, Radostin G. and {Saito}, Roberto K.},
        title = "{VIRAC2: NIR astrometry and time series photometry for 500M+ stars from the VVV and VVVX surveys}",
      journal = {\mnras},
         year = 2025,
        month = feb,
       volume = {536},
       number = {4},
        pages = {3707-3738},
          doi = {10.1093/mnras/stae2797},
archivePrefix = {arXiv},
       eprint = {2501.06295},
 primaryClass = {astro-ph.GA},
       adsurl = {https://ui.adsabs.harvard.edu/abs/2025MNRAS.536.3707S}
}

@ARTICLE{Tang2014,
       author = {{Tang}, Jing and {Bressan}, Alessandro and {Rosenfield}, Philip and {Slemer}, Alessandra and {Marigo}, Paola and {Girardi}, L{\'e}o and {Bianchi}, Luciana},
        title = "{New PARSEC evolutionary tracks of massive stars at low metallicity: testing canonical stellar evolution in nearby star-forming dwarf galaxies}",
      journal = {\mnras},
         year = 2014,
        month = dec,
       volume = {445},
       number = {4},
        pages = {4287-4305},
          doi = {10.1093/mnras/stu2029},
archivePrefix = {arXiv},
       eprint = {1410.1745},
 primaryClass = {astro-ph.SR},
       adsurl = {https://ui.adsabs.harvard.edu/abs/2014MNRAS.445.4287T}
}

@ARTICLE{Wang2007,
       author = {{Wang}, Junfeng and {Townsley}, Leisa K. and {Feigelson}, Eric D. and {Getman}, Konstantin V. and {Broos}, Patrick S. and {Garmire}, Gordon P. and {Tsujimoto}, Masahiro},
        title = "{An X-Ray Census of Young Stars in the Massive Southern Star-forming Complex NGC 6357}",
      journal = {\apjs},
         year = 2007,
        month = jan,
       volume = {168},
       number = {1},
        pages = {100-127},
          doi = {10.1086/509147},
archivePrefix = {arXiv},
       eprint = {astro-ph/0609304},
 primaryClass = {astro-ph},
       adsurl = {https://ui.adsabs.harvard.edu/abs/2007ApJS..168..100W}
}

@ARTICLE{Bolatto2013,
       author = {{Bolatto}, Alberto D. and {Wolfire}, Mark and {Leroy}, Adam K.},
        title = "{The CO-to-H$_{2}$ Conversion Factor}",
      journal = {\araa},
         year = 2013,
        month = aug,
       volume = {51},
       number = {1},
        pages = {207-268},
          doi = {10.1146/annurev-astro-082812-140944},
archivePrefix = {arXiv},
       eprint = {1301.3498},
 primaryClass = {astro-ph.GA},
       adsurl = {https://ui.adsabs.harvard.edu/abs/2013ARA&A..51..207B}
}

@ARTICLE{Chevance2020,
       author = {{Chevance}, M{\'e}lanie and {Kruijssen}, J.~M. Diederik and {Vazquez-Semadeni}, Enrique and {Nakamura}, Fumitaka and {Klessen}, Ralf and {Ballesteros-Paredes}, Javier and {Inutsuka}, Shu-ichiro and {Adamo}, Angela and {Hennebelle}, Patrick},
        title = "{The Molecular Cloud Lifecycle}",
      journal = {\ssr},
         year = 2020,
        month = apr,
       volume = {216},
       number = {4},
          eid = {50},
        pages = {50},
          doi = {10.1007/s11214-020-00674-x},
archivePrefix = {arXiv},
       eprint = {2004.06113},
 primaryClass = {astro-ph.GA},
       adsurl = {https://ui.adsabs.harvard.edu/abs/2020SSRv..216...50C}
}

@ARTICLE{Fukui2018,
       author = {{Fukui}, Yasuo and {Kohno}, Mikito and {Yokoyama}, Keiko and {Torii}, Kazufumi and {Hattori}, Yusuke and {Sano}, Hidetoshi and {Nishimura}, Atsushi and {Ohama}, Akio and {Yamamoto}, Hiroaki and {Tachihara}, Kengo},
        title = "{Molecular clouds in the NGC 6334 and NGC 6357 region: Evidence for a 100 pc-scale cloud-cloud collision triggering the Galactic mini-starbursts}",
      journal = {\pasj},
         year = 2018,
        month = may,
       volume = {70},
          eid = {S41},
        pages = {S41},
          doi = {10.1093/pasj/psy017},
archivePrefix = {arXiv},
       eprint = {1706.05771},
 primaryClass = {astro-ph.GA},
       adsurl = {https://ui.adsabs.harvard.edu/abs/2018PASJ...70S..41F}
}

@ARTICLE{Fukui2021,
       author = {{Fukui}, Yasuo and {Habe}, Asao and {Inoue}, Tsuyoshi and {Enokiya}, Rei and {Tachihara}, Kengo},
        title = "{Cloud-cloud collisions and triggered star formation}",
      journal = {\pasj},
         year = 2021,
        month = jan,
       volume = {73},
        pages = {S1-S34},
          doi = {10.1093/pasj/psaa103},
archivePrefix = {arXiv},
       eprint = {2009.05077},
 primaryClass = {astro-ph.GA},
       adsurl = {https://ui.adsabs.harvard.edu/abs/2021PASJ...73S...1F}
}

@ARTICLE{Getman2019,
       author = {{Getman}, K.~V. and {Feigelson}, E.~D. and {Kuhn}, M.~A. and {Garmire}, G.~P.},
        title = "{Gaia stellar kinematics in the head of the Orion A cloud: runaway stellar groups and gravitational infall}",
      journal = {\mnras},
         year = 2019,
        month = aug,
       volume = {487},
       number = {3},
        pages = {2977-3000},
          doi = {10.1093/mnras/stz1457},
archivePrefix = {arXiv},
       eprint = {1905.10251},
 primaryClass = {astro-ph.GA},
       adsurl = {https://ui.adsabs.harvard.edu/abs/2019MNRAS.487.2977G}
}

@ARTICLE{Getman2022,
       author = {{Getman}, Konstantin V. and {Feigelson}, Eric D. and {Garmire}, Gordon P. and {Broos}, Patrick S. and {Kuhn}, Michael A. and {Preibisch}, Thomas and {Airapetian}, Vladimir S.},
        title = "{Evolution of X-Ray Activity in <25 Myr Old Pre-main Sequence Stars}",
      journal = {\apj},
         year = 2022,
        month = aug,
       volume = {935},
       number = {1},
          eid = {43},
        pages = {43},
          doi = {10.3847/1538-4357/ac7c69},
archivePrefix = {arXiv},
       eprint = {2203.02047},
 primaryClass = {astro-ph.SR},
       adsurl = {https://ui.adsabs.harvard.edu/abs/2022ApJ...935...43G}
}

@ARTICLE{Habe1992,
       author = {{Habe}, Asao and {Ohta}, Kanji},
        title = "{Gravitational Instability Induced by a Cloud-Cloud Collision: The Case of Head-on Collisions between Clouds with Different Sizes and Densities}",
      journal = {\pasj},
         year = 1992,
        month = jul,
       volume = {44},
       number = {3},
        pages = {203-226},
          doi = {10.1093/pasj/44.3.203},
       adsurl = {https://ui.adsabs.harvard.edu/abs/1992PASJ...44..203H}
}

@ARTICLE{Haworth2015,
       author = {{Haworth}, T.~J. and {Tasker}, E.~J. and {Fukui}, Y. and {Torii}, K. and {Dale}, J.~E. and {Shima}, K. and {Takahira}, K. and {Habe}, A. and {Hasegawa}, K.},
        title = "{Isolating signatures of major cloud-cloud collisions using position-velocity diagrams}",
      journal = {\mnras},
         year = 2015,
        month = jun,
       volume = {450},
       number = {1},
        pages = {10-20},
          doi = {10.1093/mnras/stv639},
archivePrefix = {arXiv},
       eprint = {1503.06795},
 primaryClass = {astro-ph.GA},
       adsurl = {https://ui.adsabs.harvard.edu/abs/2015MNRAS.450...10H}
}

@ARTICLE{InoueFukui2013,
       author = {{Inoue}, Tsuyoshi and {Fukui}, Yasuo},
        title = "{Formation of Massive Molecular Cloud Cores by Cloud-Cloud Collision}",
      journal = {\apjl},
         year = 2013,
        month = sep,
       volume = {774},
       number = {2},
          eid = {L31},
        pages = {L31},
          doi = {10.1088/2041-8205/774/2/L31},
archivePrefix = {arXiv},
       eprint = {1305.4655},
 primaryClass = {astro-ph.GA},
       adsurl = {https://ui.adsabs.harvard.edu/abs/2013ApJ...774L..31I}
}

@ARTICLE{Kawamura2009,
       author = {{Kawamura}, Akiko and {Mizuno}, Yoji and {Minamidani}, Tetsuhiro and {Filipovi{\'c}}, Miroslav D. and {Staveley-Smith}, Lister and {Kim}, Sungeun and {Mizuno}, Norikazu and {Onishi}, Toshikazu and {Mizuno}, Akira and {Fukui}, Yasuo},
        title = "{The Second Survey of the Molecular Clouds in the Large Magellanic Cloud by NANTEN. II. Star Formation}",
      journal = {\apjs},
         year = 2009,
        month = sep,
       volume = {184},
       number = {1},
        pages = {1-17},
          doi = {10.1088/0067-0049/184/1/1},
archivePrefix = {arXiv},
       eprint = {0908.1168},
 primaryClass = {astro-ph.CO},
       adsurl = {https://ui.adsabs.harvard.edu/abs/2009ApJS..184....1K}
}

@ARTICLE{LadaLada2003,
       author = {{Lada}, Charles J. and {Lada}, Elizabeth A.},
        title = "{Embedded Clusters in Molecular Clouds}",
      journal = {\araa},
         year = 2003,
        month = jan,
       volume = {41},
        pages = {57-115},
          doi = {10.1146/annurev.astro.41.011802.094844},
archivePrefix = {arXiv},
       eprint = {astro-ph/0301540},
 primaryClass = {astro-ph},
       adsurl = {https://ui.adsabs.harvard.edu/abs/2003ARA&A..41...57L}
}

@ARTICLE{MangumShirley2015,
       author = {{Mangum}, Jeffrey G. and {Shirley}, Yancy L.},
        title = "{How to Calculate Molecular Column Density}",
      journal = {\pasp},
         year = 2015,
        month = mar,
       volume = {127},
       number = {949},
        pages = {266},
          doi = {10.1086/680323},
archivePrefix = {arXiv},
       eprint = {1501.01703},
 primaryClass = {astro-ph.IM},
       adsurl = {https://ui.adsabs.harvard.edu/abs/2015PASP..127..266M}
}

@ARTICLE{OrdenesHuanca2024,
       author = {{Ordenes-Huanca}, C. and {Zoccali}, M. and {Bayo}, A. and {Cuadra}, J. and {Contreras Ramos}, R. and {Rojas-Arriagada}, A.},
        title = "{Unveiling the structural content of NGC 6357 via kinematics and NIR variability}",
      journal = {\mnras},
         year = 2024,
        month = sep,
       volume = {533},
       number = {1},
        pages = {841-859},
          doi = {10.1093/mnras/stae1862},
archivePrefix = {arXiv},
       eprint = {2407.17577},
 primaryClass = {astro-ph.SR},
       adsurl = {https://ui.adsabs.harvard.edu/abs/2024MNRAS.533..841O}
}

@ARTICLE{Russeil2016,
       author = {{Russeil}, D. and {Tig{\'e}}, J. and {Adami}, C. and {Anderson}, L.~D. and {Schneider}, N. and {Zavagno}, A. and {Samal}, M.~R. and {Amram}, P. and {Guennou}, L. and {Le Coarer}, E. and {Walsh}, A. and {Longmore}, S.~N. and {Purcell}, C.},
        title = "{NGC 6334 and NGC 6357: H{\ensuremath{\alpha}} kinematics and the nature of the H II regions}",
      journal = {\aap},
         year = 2016,
        month = mar,
       volume = {587},
          eid = {A135},
        pages = {A135},
          doi = {10.1051/0004-6361/201424484},
       adsurl = {https://ui.adsabs.harvard.edu/abs/2016A&A...587A.135R}
}

@ARTICLE{Takahira2014,
       author = {{Takahira}, Ken and {Tasker}, Elizabeth J. and {Habe}, Asao},
        title = "{Do Cloud-Cloud Collisions Trigger High-mass Star Formation? I. Small Cloud Collisions}",
      journal = {\apj},
         year = 2014,
        month = sep,
       volume = {792},
       number = {1},
          eid = {63},
        pages = {63},
          doi = {10.1088/0004-637X/792/1/63},
archivePrefix = {arXiv},
       eprint = {1407.4544},
 primaryClass = {astro-ph.GA},
       adsurl = {https://ui.adsabs.harvard.edu/abs/2014ApJ...792...63T}
}

@ARTICLE{VasilievBaumgardt2021,
       author = {{Vasiliev}, Eugene and {Baumgardt}, Holger},
        title = "{Gaia EDR3 view on galactic globular clusters}",
      journal = {\mnras},
         year = 2021,
        month = aug,
       volume = {505},
       number = {4},
        pages = {5978-6002},
          doi = {10.1093/mnras/stab1475},
archivePrefix = {arXiv},
       eprint = {2102.09568},
 primaryClass = {astro-ph.GA},
       adsurl = {https://ui.adsabs.harvard.edu/abs/2021MNRAS.505.5978V}
}

@ARTICLE{Zhou2025,
       author = {{Zhou}, Ji-Xuan and {Li}, Guang-Xing and {Chen}, Bing-Qiu},
        title = "{Kinematics of the interstellar medium using Gaia: A catalogue of 102 young stellar object─molecular cloud associations within 3.5 kpc of the Sun with 3D velocities}",
      journal = {\mnras},
         year = 2025,
        month = sep,
       volume = {542},
       number = {1},
        pages = {52-61},
          doi = {10.1093/mnras/staf1222},
       adsurl = {https://ui.adsabs.harvard.edu/abs/2025MNRAS.542...52Z}
}

@INPROCEEDINGS{CastelliKurucz2003,
       author = {{Castelli}, F. and {Kurucz}, R.~L.},
        title = "{New Grids of ATLAS9 Model Atmospheres}",
    booktitle = {Modelling of Stellar Atmospheres},
         year = 2003,
       editor = {{Piskunov}, N. and {Weiss}, W.~W. and {Gray}, D.~F.},
       series = {IAU Symposium},
       volume = {210},
        month = jan,
        pages = {A20},
          doi = {10.48550/arXiv.astro-ph/0405087},
archivePrefix = {arXiv},
       eprint = {astro-ph/0405087},
 primaryClass = {astro-ph},
       adsurl = {https://ui.adsabs.harvard.edu/abs/2003IAUS..210P.A20C}
}

@ARTICLE{Churchwell2009,
       author = {{Churchwell}, Ed and {Babler}, Brian L. and {Meade}, Marilyn R. and {Whitney}, Barbara A. and {Benjamin}, Robert and {Indebetouw}, Remy and {Cyganowski}, Claudia and {Robitaille}, Thomas P. and {Povich}, Matthew and {Watson}, Christer and {Bracker}, Steve},
        title = "{The Spitzer/GLIMPSE Surveys: A New View of the Milky Way}",
      journal = {\pasp},
         year = 2009,
        month = mar,
       volume = {121},
       number = {877},
        pages = {213},
          doi = {10.1086/597811},
       adsurl = {https://ui.adsabs.harvard.edu/abs/2009PASP..121..213C}
}

@ARTICLE{Astropy2022,
       author = {{Astropy Collaboration} and {Price-Whelan}, Adrian M. and {Lim}, Pey Lian and {Earl}, Nicholas and {Starkman}, Nathaniel and {Bradley}, Larry and {Shupe}, David L. and {Patil}, Aarya A. and {Corrales}, Lia and {Brasseur}, C.~E. and {N{\"o}the}, Maximilian and {Donath}, Axel and {Tollerud}, Erik and {Morris}, Brett M. and {Ginsburg}, Adam and {Vaher}, Eero and {Weaver}, Benjamin A. and {Tocknell}, James and {Jamieson}, William and {van Kerkwijk}, Marten H. and {Robitaille}, Thomas P. and {Merry}, Bruce and {Bachetti}, Matteo and {G{\"u}nther}, H. Moritz and {Aldcroft}, Thomas L. and {Alvarado-Montes}, Jaime A. and {Archibald}, Anne M. and {B{\'o}di}, Attila and {Bapat}, Shreyas and {Barentsen}, Geert and {Baz{\'a}n}, Juanjo and {Biswas}, Manish and {Boquien}, M{\'e}d{\'e}ric and {Burke}, D.~J. and {Cara}, Daria and {Cara}, Mihai and {Conroy}, Kyle E. and {Conseil}, Simon and {Craig}, Matthew W. and {Cross}, Robert M. and {Cruz}, Kelle L. and {D'Eugenio}, Francesco and {Dencheva}, Nadia and {Devillepoix}, Hadrien A.~R. and {Dietrich}, J{\"o}rg P. and {Eigenbrot}, Arthur Davis and {Erben}, Thomas and {Ferreira}, Leonardo and {Foreman-Mackey}, Daniel and {Fox}, Ryan and {Freij}, Nabil and {Garg}, Suyog and {Geda}, Robel and {Glattly}, Lauren and {Gondhalekar}, Yash and {Gordon}, Karl D. and {Grant}, David and {Greenfield}, Perry and {Groener}, Austen M. and {Guest}, Steve and {Gurovich}, Sebastian and {Handberg}, Rasmus and {Hart}, Akeem and {Hatfield-Dodds}, Zac and {Homeier}, Derek and {Hosseinzadeh}, Griffin and {Jenness}, Tim and {Jones}, Craig K. and {Joseph}, Prajwel and {Kalmbach}, J. Bryce and {Karamehmetoglu}, Emir and {Ka{\l}uszy{\'n}ski}, Miko{\l}aj and {Kelley}, Michael S.~P. and {Kern}, Nicholas and {Kerzendorf}, Wolfgang E. and {Koch}, Eric W. and {Kulumani}, Shankar and {Lee}, Antony and {Ly}, Chun and {Ma}, Zhiyuan and {MacBride}, Conor and {Maljaars}, Jakob M. and {Muna}, Demitri and {Murphy}, N.~A. and {Norman}, Henrik and {O'Steen}, Richard and {Oman}, Kyle A. and {Pacifici}, Camilla and {Pascual}, Sergio and {Pascual-Granado}, J. and {Patil}, Rohit R. and {Perren}, Gabriel I. and {Pickering}, Timothy E. and {Rastogi}, Tanuj and {Roulston}, Benjamin R. and {Ryan}, Daniel F. and {Rykoff}, Eli S. and {Sabater}, Jose and {Sakurikar}, Parikshit and {Salgado}, Jes{\'u}s and {Sanghi}, Aniket and {Saunders}, Nicholas and {Savchenko}, Volodymyr and {Schwardt}, Ludwig and {Seifert-Eckert}, Michael and {Shih}, Albert Y. and {Jain}, Anany Shrey and {Shukla}, Gyanendra and {Sick}, Jonathan and {Simpson}, Chris and {Singanamalla}, Sudheesh and {Singer}, Leo P. and {Singhal}, Jaladh and {Sinha}, Manodeep and {Sip{\H{o}}cz}, Brigitta M. and {Spitler}, Lee R. and {Stansby}, David and {Streicher}, Ole and {{\v{S}}umak}, Jani and {Swinbank}, John D. and {Taranu}, Dan S. and {Tewary}, Nikita and {Tremblay}, Grant R. and {de Val-Borro}, Miguel and {Van Kooten}, Samuel J. and {Vasovi{\'c}}, Zlatan and {Verma}, Shresth and {de Miranda Cardoso}, Jos{\'e} Vin{\'\i}cius and {Williams}, Peter K.~G. and {Wilson}, Tom J. and {Winkel}, Benjamin and {Wood-Vasey}, W.~M. and {Xue}, Rui and {Yoachim}, Peter and {Zhang}, Chen and {Zonca}, Andrea and {Astropy Project Contributors}},
        title = "{The Astropy Project: Sustaining and Growing a Community-oriented Open-source Project and the Latest Major Release (v5.0) of the Core Package}",
      journal = {\apj},
         year = 2022,
        month = aug,
       volume = {935},
       number = {2},
          eid = {167},
        pages = {167},
          doi = {10.3847/1538-4357/ac7c74},
archivePrefix = {arXiv},
       eprint = {2206.14220},
 primaryClass = {astro-ph.IM},
       adsurl = {https://ui.adsabs.harvard.edu/abs/2022ApJ...935..167A}
}

@ARTICLE{Harris2020,
       author = {{Harris}, Charles R. and {Millman}, K. Jarrod and {van der Walt}, St{\'e}fan J. and {Gommers}, Ralf and {Virtanen}, Pauli and {Cournapeau}, David and {Wieser}, Eric and {Taylor}, Julian and {Berg}, Sebastian and {Smith}, Nathaniel J. and {Kern}, Robert and {Picus}, Matti and {Hoyer}, Stephan and {van Kerkwijk}, Marten H. and {Brett}, Matthew and {Haldane}, Allan and {del R{\'\i}o}, Jaime Fern{\'a}ndez and {Wiebe}, Mark and {Peterson}, Pearu and {G{\'e}rard-Marchant}, Pierre and {Sheppard}, Kevin and {Reddy}, Tyler and {Weckesser}, Warren and {Abbasi}, Hameer and {Gohlke}, Christoph and {Oliphant}, Travis E.},
        title = "{Array programming with NumPy}",
      journal = {\nat},
         year = 2020,
        month = sep,
       volume = {585},
       number = {7825},
        pages = {357-362},
          doi = {10.1038/s41586-020-2649-2},
archivePrefix = {arXiv},
       eprint = {2006.10256},
 primaryClass = {cs.MS},
       adsurl = {https://ui.adsabs.harvard.edu/abs/2020Natur.585..357H}
}

@ARTICLE{Virtanen2020,
       author = {{Virtanen}, Pauli and {Gommers}, Ralf and {Oliphant}, Travis E. and {Haberland}, Matt and {Reddy}, Tyler and {Cournapeau}, David and {Burovski}, Evgeni and {Peterson}, Pearu and {Weckesser}, Warren and {Bright}, Jonathan and {van der Walt}, St{\'e}fan J. and {Brett}, Matthew and {Wilson}, Joshua and {Millman}, K. Jarrod and {Mayorov}, Nikolay and {Nelson}, Andrew R.~J. and {Jones}, Eric and {Kern}, Robert and {Larson}, Eric and {Carey}, C.~J. and {Polat}, {\.I}lhan and {Feng}, Yu and {Moore}, Eric W. and {VanderPlas}, Jake and {Laxalde}, Denis and {Perktold}, Josef and {Cimrman}, Robert and {Henriksen}, Ian and {Quintero}, E.~A. and {Harris}, Charles R. and {Archibald}, Anne M. and {Ribeiro}, Ant{\^o}nio H. and {Pedregosa}, Fabian and {van Mulbregt}, Paul and {SciPy 1.  0 Contributors}},
        title = "{SciPy 1.0: fundamental algorithms for scientific computing in Python}",
      journal = {Nature Methods},
         year = 2020,
        month = feb,
       volume = {17},
        pages = {261-272},
          doi = {10.1038/s41592-019-0686-2},
archivePrefix = {arXiv},
       eprint = {1907.10121},
 primaryClass = {cs.MS},
       adsurl = {https://ui.adsabs.harvard.edu/abs/2020NaMet..17..261V}
}

@misc{McKinney2010,
       author = {{McKinney}, Wes},
        title = "{Data Structures for Statistical Computing in Python}",
 howpublished = {Proceedings of the 9th Python in Science Conference},
         year = 2010,
        month = jan,
          doi = {10.25080/Majora-92bf1922-00a},
       adsurl = {https://ui.adsabs.harvard.edu/abs/2010scpy.soft.....M}
}

@ARTICLE{Hunter2007,
       author = {{Hunter}, John D.},
        title = "{Matplotlib: A 2D Graphics Environment}",
      journal = {Computing in Science and Engineering},
         year = 2007,
        month = jan,
       volume = {9},
       number = {3},
        pages = {90-95},
          doi = {10.1109/MCSE.2007.55},
       adsurl = {https://ui.adsabs.harvard.edu/abs/2007CSE.....9...90H}
}

@ARTICLE{Ginsburg2019,
       author = {{Ginsburg}, Adam and {Sip{\H{o}}cz}, Brigitta M. and {Brasseur}, C.~E. and {Cowperthwaite}, Philip S. and {Craig}, Matthew W. and {Deil}, Christoph and {Guillochon}, James and {Guzman}, Giannina and {Liedtke}, Simon and {Lian Lim}, Pey and {Lockhart}, Kelly E. and {Mommert}, Michael and {Morris}, Brett M. and {Norman}, Henrik and {Parikh}, Madhura and {Persson}, Magnus V. and {Robitaille}, Thomas P. and {Segovia}, Juan-Carlos and {Singer}, Leo P. and {Tollerud}, Erik J. and {de Val-Borro}, Miguel and {Valtchanov}, Ivan and {Woillez}, Julien and {Astroquery Collaboration} and {a subset of astropy Collaboration}},
        title = "{astroquery: An Astronomical Web-querying Package in Python}",
      journal = {\aj},
         year = 2019,
        month = mar,
       volume = {157},
       number = {3},
          eid = {98},
        pages = {98},
          doi = {10.3847/1538-3881/aafc33},
archivePrefix = {arXiv},
       eprint = {1901.04520},
 primaryClass = {astro-ph.IM},
       adsurl = {https://ui.adsabs.harvard.edu/abs/2019AJ....157...98G}
}

@ARTICLE{Pedregosa2011,
       author = {{Pedregosa}, Fabian and {Varoquaux}, Ga{\"e}l and {Gramfort}, Alexandre and {Michel}, Vincent and {Thirion}, Bertrand and {Grisel}, Olivier and {Blondel}, Mathieu and {M{\"u}ller}, Andreas and {Nothman}, Joel and {Louppe}, Gilles and {Prettenhofer}, Peter and {Weiss}, Ron and {Dubourg}, Vincent and {Vanderplas}, Jake and {Passos}, Alexandre and {Cournapeau}, David and {Brucher}, Matthieu and {Perrot}, Matthieu and {Duchesnay}, {\'E}douard},
        title = "{Scikit-learn: Machine Learning in Python}",
      journal = {Journal of Machine Learning Research},
         year = 2011,
        month = oct,
       volume = {12},
        pages = {2825-2830},
          doi = {10.48550/arXiv.1201.0490},
archivePrefix = {arXiv},
       eprint = {1201.0490},
 primaryClass = {cs.LG},
       adsurl = {https://ui.adsabs.harvard.edu/abs/2011JMLR...12.2825P}
}

@ARTICLE{2016A&A...595A...1G,
       author = {{Gaia Collaboration} and {Prusti}, T. and {de Bruijne}, J.~H.~J. and {Brown}, A.~G.~A. and {Vallenari}, A. and {Babusiaux}, C. and {Bailer-Jones}, C.~A.~L. and {Bastian}, U. and {Biermann}, M. and {Evans}, D.~W. and {Eyer}, L. and {Jansen}, F. and {Jordi}, C. and {Klioner}, S.~A. and {Lammers}, U. and {Lindegren}, L. and {Luri}, X. and {Mignard}, F. and {Milligan}, D.~J. and {Panem}, C. and {Poinsignon}, V. and {Pourbaix}, D. and {Randich}, S. and {Sarri}, G. and {Sartoretti}, P. and {Siddiqui}, H.~I. and {Soubiran}, C. and {Valette}, V. and {van Leeuwen}, F. and {Walton}, N.~A. and {Aerts}, C. and {Arenou}, F. and {Cropper}, M. and {Drimmel}, R. and {H{\o}g}, E. and {Katz}, D. and {Lattanzi}, M.~G. and {O'Mullane}, W. and {Grebel}, E.~K. and {Holland}, A.~D. and {Huc}, C. and {Passot}, X. and {Bramante}, L. and {Cacciari}, C. and {Casta{\~n}eda}, J. and {Chaoul}, L. and {Cheek}, N. and {De Angeli}, F. and {Fabricius}, C. and {Guerra}, R. and {Hern{\'a}ndez}, J. and {Jean-Antoine-Piccolo}, A. and {Masana}, E. and {Messineo}, R. and {Mowlavi}, N. and {Nienartowicz}, K. and {Ord{\'o}{\~n}ez-Blanco}, D. and {Panuzzo}, P. and {Portell}, J. and {Richards}, P.~J. and {Riello}, M. and {Seabroke}, G.~M. and {Tanga}, P. and {Th{\'e}venin}, F. and {Torra}, J. and {Els}, S.~G. and {Gracia-Abril}, G. and {Comoretto}, G. and {Garcia-Reinaldos}, M. and {Lock}, T. and {Mercier}, E. and {Altmann}, M. and {Andrae}, R. and {Astraatmadja}, T.~L. and {Bellas-Velidis}, I. and {Benson}, K. and {Berthier}, J. and {Blomme}, R. and {Busso}, G. and {Carry}, B. and {Cellino}, A. and {Clementini}, G. and {Cowell}, S. and {Creevey}, O. and {Cuypers}, J. and {Davidson}, M. and {De Ridder}, J. and {de Torres}, A. and {Delchambre}, L. and {Dell'Oro}, A. and {Ducourant}, C. and {Fr{\'e}mat}, Y. and {Garc{\'\i}a-Torres}, M. and {Gosset}, E. and {Halbwachs}, J.-L. and {Hambly}, N.~C. and {Harrison}, D.~L. and {Hauser}, M. and {Hestroffer}, D. and {Hodgkin}, S.~T. and {Huckle}, H.~E. and {Hutton}, A. and {Jasniewicz}, G. and {Jordan}, S. and {Kontizas}, M. and {Korn}, A.~J. and {Lanzafame}, A.~C. and {Manteiga}, M. and {Moitinho}, A. and {Muinonen}, K. and {Osinde}, J. and {Pancino}, E. and {Pauwels}, T. and {Petit}, J.-M. and {Recio-Blanco}, A. and {Robin}, A.~C. and {Sarro}, L.~M. and {Siopis}, C. and {Smith}, M. and {Smith}, K.~W. and {Sozzetti}, A. and {Thuillot}, W. and {van Reeven}, W. and {Viala}, Y. and {Abbas}, U. and {Abreu Aramburu}, A. and {Accart}, S. and {Aguado}, J.~J. and {Allan}, P.~M. and {Allasia}, W. and {Altavilla}, G. and {{\'A}lvarez}, M.~A. and {Alves}, J. and {Anderson}, R.~I. and {Andrei}, A.~H. and {Anglada Varela}, E. and {Antiche}, E. and {Antoja}, T. and {Ant{\'o}n}, S. and {Arcay}, B. and {Atzei}, A. and {Ayache}, L. and {Bach}, N. and {Baker}, S.~G. and {Balaguer-N{\'u}{\~n}ez}, L. and {Barache}, C. and {Barata}, C. and {Barbier}, A. and {Barblan}, F. and {Baroni}, M. and {Barrado y Navascu{\'e}s}, D. and {Barros}, M. and {Barstow}, M.~A. and {Becciani}, U. and {Bellazzini}, M. and {Bellei}, G. and {Bello Garc{\'\i}a}, A. and {Belokurov}, V. and {Bendjoya}, P. and {Berihuete}, A. and {Bianchi}, L. and {Bienaym{\'e}}, O. and {Billebaud}, F. and {Blagorodnova}, N. and {Blanco-Cuaresma}, S. and {Boch}, T. and {Bombrun}, A. and {Borrachero}, R. and {Bouquillon}, S. and {Bourda}, G. and {Bouy}, H. and {Bragaglia}, A. and {Breddels}, M.~A. and {Brouillet}, N. and {Br{\"u}semeister}, T. and {Bucciarelli}, B. and {Budnik}, F. and {Burgess}, P. and {Burgon}, R. and {Burlacu}, A. and {Busonero}, D. and {Buzzi}, R. and {Caffau}, E. and {Cambras}, J. and {Campbell}, H. and {Cancelliere}, R. and {Cantat-Gaudin}, T. and {Carlucci}, T. and {Carrasco}, J.~M. and {Castellani}, M. and {Charlot}, P. and {Charnas}, J. and {Charvet}, P. and {Chassat}, F. and {Chiavassa}, A. and {Clotet}, M. and {Cocozza}, G. and {Collins}, R.~S. and {Collins}, P. and {Costigan}, G.},
        title = "{The Gaia mission}",
      journal = {\aap},
         year = 2016,
        month = nov,
       volume = {595},
          eid = {A1},
        pages = {A1},
          doi = {10.1051/0004-6361/201629272},
archivePrefix = {arXiv},
       eprint = {1609.04153},
 primaryClass = {astro-ph.IM},
       adsurl = {https://ui.adsabs.harvard.edu/abs/2016A&A...595A...1G}
}

\end{document}